\documentclass[twocolumn,superscriptaddress,showpacs,preprintnumbers,amsmath,amssymb,floatfix]{revtex4}

\usepackage{graphicx,amssymb}
\usepackage{amsmath,latexsym}
\usepackage{amsfonts,wasysym}
\usepackage[utf8]{inputenc}
\usepackage{dsfont}
\usepackage{longtable}
\usepackage[normalem]{ulem}

\newlength{\barheight}
\newlength{\stayunder}

\begin{document}

\title{Zero temperature phase diagram of the square-shoulder system}

\author{Gernot J. Pauschenwein} 
\affiliation{} 

\author{Gerhard Kahl}
\affiliation{Institut f\"ur Theoretische Physik and Center for 
Computational Materials Science (CMS), Technische Universit\"at Wien, 
Wiedner Hauptstra{\ss}e 8-10, A-1040 Wien, Austria}

\date{\today}

\begin{abstract}
Particles that interact via a square-shoulder potential, consisting of
an impenetrable hard core with an adjacent, repulsive, step-like
corona, are able to self-organize in a surprisingly rich variety of
rather unconventional ordered structures. Using optimization
strategies that are based on ideas of genetic algorithms we encounter,
as we systematically increase the pressure, the following archetypes
of aggregates: low-symmetry cluster and columnar phases, followed by
lamellar particle arrangements, until at high pressure values compact,
high-symmetry lattices emerge. These structures are characterized in
the NPT ensemble as configurations of minimum Gibbs free energy. Based
on simple considerations, i.e., basically minimizing the number of
overlapping coronae while maximizing at the same time the density, the
sequence of emerging structures can easily be understood.
\begin{center}
\it submitted to J. Chem. Phys.
\end{center}
\end{abstract}


\maketitle

\section{Introduction}
\label{sec:intro}

For more than 25 years considerable effort has been dedicated to study
the thermodynamic, structural, and dynamical properties of hard core
particles with an adjacent soft repulsive shoulder, i.e., so-called
core softened potentials. This class of potentials was probably first
considered by Hemmer and Stell in 1970 \cite{Hem70} in a model where
the soft repulsion was characterized by a linear ramp with an
additional, weak attractive tail. The system was introduced in an
effort to study the possibility of the occurrence of more than one
critical point in the phase diagram of a simple model system. In
numerous, subsequent investigations evidence was provided for a
surprisingly rich variety of rather unusual properties of this class
of systems: these features range from isostructural solid-solid
transitions, where possibly several solid structures are involved
(see, e.g., \cite{Kin76a,Kin76b,Ras97,Bol97,Lang00,Vel00}), over a very
complex phase diagram of the solid phases \cite{Jag98,Jag99JCP}, to
different sorts of anomalous behavior, encountered in the static
and/or in the dynamic properties (see, e.g.,
\cite{Sad98,Jag99JCPb,Yan06,Kum05,Yan05}).

Within this class of core softened potentials the square-shoulder
interaction, consisting of an impenetrable hard core with an adjacent,
repulsive shoulder (or corona), is undoubtedly the simplest
representative.  Despite its simple, radially symmetric functional
form this system is nevertheless able to offer a large variety of
unexpected features, which are mostly related to its structural
properties. This propensity and ability for unconventional
self-assembly scenarios was already discovered in the remarkable study
by Jagla \cite{Jag99JCP} on a particular family of {\it
  two}-dimensional core softened systems, where the square-shoulder
system was included as a special case: in this contribution evidence
was given that the particles are able to self-organize in a
surprisingly broad variety of highly complex ordered structures. In
subsequent work on the two-dimensional case these particle
arrangements that include, amongst others, cage- or lane-formation as
well as micellar or inverse micellar configurations were confirmed or
newly discovered both in computer simulations \cite{Mal03,Mal04} and
in theoretical investigations \cite{Gla07}. A more systematic study of
the ordered particle configurations of the two-dimensional
square-shoulder system was presented in \cite{For08Kah}.

The aim of the present contribution is to investigate in a systematic
and thorough way the ordered particle arrangements of the
square-shoulder system in {\it three} dimensions. To this end we study
the system at $T = 0$ and in the NPT ensemble; thus, we search for
configurations that minimize the Gibbs free energy, which we will term
-- to be consistent with previous contributions -- as minimum energy
configurations (MECs). To provide a deeper insight into the
self-assembly strategies of the system we have considered a small, an
intermediate, and a large shoulder range. While preliminary results
have already been presented in \cite{Pau08}, we identify in the
present, more systematic investigation an overwhelmingly rich variety
of MECs. Analyzing these data, we give evidence that these MECs can be
grouped together in four structural archetypes, that emerge in
dependence of the value of the pressure, $P$, that is exerted on the
system: {\it cluster} structures are preferentially formed at low
$P$-values, while {\it columnar} and {\it lamellar} structures are
predominantly identified at intermediate pressure values; finally,
{\it compact} particle configurations emerge at high pressure. While
this general rule might still be less obvious at small shoulder range,
it is nearly perfectly obeyed for an intermediate shoulder width and
definitely holds for the case of a broad corona. With its simple
functional form the inter-particle interaction offers not only many
computational advantages \cite{Zih01}. It allows to {\it understand}
via simple geometrical considerations the system's self-assembly
strategy: it is in particular the range of the shoulder that turns out
to be responsible in a highly sophisticated way for the formation of
the complex structures. In addition, the flat energetic plateau
of the shoulder with its finite range represents a very sensitive
antenna to distinguish between energetically competing structures. Our
observations provide a deeper insight into the system's strategy to
form ordered equilibrium particle configurations, a knowledge that
might be useful to understand self-assembly processes in other systems
with more complex interactions.

Objections against the simple functional form of the potential are
refuted by the argument that it is able to capture the essential
features of colloidal particles with core-corona architecture as they
are, for instance, treated in \cite{Nor05} and references therein. And
indeed, several of the MECs that we could identify had already been
encountered in previous theoretical, experimental, and computer
simulation investigations: micellar and inverse micellar structures
\cite{Pie06,Gla07}, spirals \cite{Cam05}, chains and layers
\cite{Jag98,Jag99JCP,Jag99JCPb,Cam03,Gla07,Can06}, and cluster phases
\cite{Str04,Mla06PRL,Mla06PRLERR}, to name a few examples.

Although the identification of MECs represents a 'simple' optimization
problem of the Gibbs free energy, its solution has turned out to be
highly non-trivial. In this contribution we present a systematic
sequence of ordered MECs for the three selected values of shoulder
width. This achievement is mainly due to our search strategy, which is
based on ideas of genetic algorithms (GA). Introduced already several
decades ago in a completely different context \cite{Hol75}, these
approaches have meanwhile become a highly appreciated optimization
tool to identify ordered particle arrangements both in hard
\cite{Woo99,Oga06,Oga08} as well as in soft matter systems
\cite{Got04,Got05Kah,Got05Lik,For08Kah,For08LoV,Pau08}. The high
reliability, flexibility, and efficiency of GA-based optimization
strategies, in combination with a particular search strategy that is
intimately related to the simple functional form of the
square-shoulder potential, make us believe that the sequences of MECs,
that will be presented and discussed in the following, are complete.

The paper is organized as follows. In section \ref{sec:model} we
briefly present the square-shoulder system. The subsequent section
deals with the theoretical tools of this contribution: the GA-based
search strategy as well as the theoretical considerations to identify
close-packed particle arrangements of the system as a function of the
shoulder width. The results of our investigations are summarized in
section \ref{sec:results}: we start with the close-packed particle
arrangements (as they play a key-role in the search strategy) and
present and discuss in the following the MECs that we have identified
for the three different cases of shoulder width. The conclusions of
the contribution are summarized in section \ref{sec:conclu}, which
also contain{\sf\sout{s}} the discussion of possible future work.

\section{Model}
\label{sec:model}

We consider a system of particles that interact via the
square-shoulder potential, which we parameterize as follows:
\begin{equation}
\Phi(r) = \left\{ \begin{array} {l@{~~~~~~~~}l} 
                  \infty & r \le \sigma \\ 
                  \epsilon & \sigma < r \le \lambda \sigma \\
                  0 & \lambda \sigma < r  \\
                             \end{array}
                             \right. .
\end{equation}
$\sigma$ is the diameter of the impenetrable core and $\lambda \sigma$
is the width of the adjacent, repulsive shoulder (or corona) of height
$\epsilon$, $\epsilon > 0$. Further, we introduce the number-density,
$\rho = N/V$ ($N$ being the number of particles and $V$ being the
volume of the system), and the dimensionless number-density,
$\rho^\star = \rho \sigma^3$. Thermodynamic quantities will be used in
the following reduced units: pressure, $P^\star = P
\sigma^3/\epsilon$, internal energy, $e^\star = E/(N\epsilon)$, and
Gibbs free energy, $g^\star = G/(N\epsilon)$. Since we perform our
investigations at $T = 0$, $G = E + PV$ and hence $g^\star = e^\star +
P^\star/\rho^\star$.

The simple functional form of the square-shoulder potential with its
constant potential barrier and its finite range makes the system the
'quintessential test system' \cite{Zih01} for the purpose of the
present contribution. It also simplifies considerably thermodynamic
considerations. For a given periodic particle arrangement, which we
characterize by the number of overlapping coronae, $e^\star$ is a
rational number: it is given as the ratio of the number of overlaps
per particle in the unit cell, divided by the number of these
particles, which we denote as $n_b$. For this particle arrangement
$g^\star = e^\star + P^\star/\rho^\star$, is therefore a linear
function of the pressure, $P^\star$, and is consequently represented
in the $(g^\star, P^\star)$-plane by a straight line: its slope is
given by $1/\rho^\star$, while its intercept is the energy of the
configuration, $e^\star$. The limiting particle arrangement at low
pressure is easily identified as a close-packed arrangement of spheres
with diameter $\lambda \sigma$: thus $e^\star = 0$ and the slope of
$g^\star$ in the $(g^\star, P^\star)$-diagram is given by
$1/\rho^\star_{\rm min} = \lambda^3/\sqrt 2$. For the high pressure
limit the situation is more delicate: while the slope of $g^\star$ as
a function of $P^\star$ is easily identified for obvious reasons as
$1/\rho^\star_{\rm max} = 1/\sqrt 2$, the value of $e^\star$ depends
in a sensitive way on $\lambda$. In subsection \ref{subsec:th_cp} we
will give evidence that the square-shoulder system shows a rich
variety of close-packed scenarios as $\lambda$ varies.

With the above considerations in mind, we can anticipate that the
$g^\star$-values of all MECs will be located on a sequence of
intersecting straight lines in the $(g^\star, P^\star)$-plane, each of
them being characterized by a slope of $1/\rho^\star$, with
$1/\rho^\star_{\rm min} > 1/\rho^\star > 1/\rho^\star_{\rm max}$, and
by an intercept $e^\star$. This fact will simplify considerably our
search for MECs (see discussion in subsection \ref{subsec:res_mecs}).

\section{Theory}
\label{sec:theory}

\subsection{Genetic algorithms}
\label{subsec:ga}

The MECs of our system have been identified with a search strategy
that is based on ideas of genetic algorithms (GAs). GAs are very
general optimization tools that model natural evolution processes,
such as recombination, mutation, or survival of the fittest
\cite{Hol75}. Their successful applications in a wide range of fields
demonstrate their flexibility and reliability. The basic ideas of GAs
can be summarized as follows: the central quantity of this concept is
an individual, ${\cal I}$, which represents a possible solution to the
problem.  The quality of a solution, i.e., of an individual ${\cal
  I}$, is measured via a so-called fitness function, $f({\cal
  I})$. Individuals with a higher fitness value are assumed to be of
higher quality. In our search for ordered particle configurations that
minimize the Gibbs free energy, an individual corresponds to a lattice
while the fitness function is related to $G$ and will be specified
below. Starting from a large number of individuals, which represent
the initial generation, individuals of a subsequent generation are
created with recombination and mutation processes, both of them having
a highly stochastic character. Individuals with a higher fitness value
are preferred in the reproduction process. In addition, mutation
operations are performed on the individuals with some probability
$p_{\rm m}$. By iterating this process we create a reasonably large
number of generations. The final result of the GA-based search
strategy is the individual with the overall highest fitness value.

For our particular problem an individual ${\cal I}$ is identified by a
(possibly non-simple) periodic crystal structure. Due to the highly
stochastic character of the reproduction and of the mutation
processes, a straightforward implementation of the algorithm is prone
to propose a large number of ordered configurations where the hard
cores of the particles overlap and which therefore correspond to
unphysical particle arrangements. This, in turn, causes a drastic
reduction of the efficiency of the algorithms. To overcome this
problem we have developed a particular parameterization of an {\it
  arbitrary} simple lattice via three lattice vectors $\{ {\bf a}_1,
{\bf a}_2, {\bf a}_3 \}$ \cite{Pau08mindist}: here, $a_1 = | {\bf
  a}_1|$ represents the shortest possible distance between two lattice
sites in the entire lattice and $a_2 = | {\bf a}_2|$ is the second
smallest distance in the lattice (i.e., $a_1 \le a_2$) with ${\bf
  a}_1$ and ${\bf a}_2$ being linearly independent; finally, a similar
relation holds between ${\bf a}_3$ on one side and ${\bf a}_1$ and
${\bf a}_2$ on the other side. Thus if $a_1 > \sigma$, it is
guaranteed that the hard cores of the particles will not overlap and
consequently the GA will create only simple lattices where overlap of
the cores is avoided {\it a priori}. For non simple lattices, the
distances between all particles within the unit cell and including
also the particles of the 26 neighboring cells have to be
determined. If the smallest of these distances, $l_0$, is smaller than
$a_1$, then the lattice is scaled with a factor $a_1/l_0$. The rather
complex formalism is most conveniently implemented in the NPT
ensemble. Thus, a state is characterized by a value for the pressure,
$P$, while the equilibrium density, $\rho$, is a result of the
optimization procedure. For details we refer to \cite{Pau08mindist}.

For the implementation of the individuals we use the encoding strategy
presented in \cite{Got05Lik}. Distances and angles are encoded in
binary strings with a length ranging from four to six. Since in the
NPT ensemble the Gibbs free energy has to be minimized, we use the
following fitness function

\begin{equation}
f({\cal I}) = \exp\{ -[G({\cal I}) - G({\cal I}_0)]/G({\cal I}_0)\} ,
\end{equation}
where ${\cal I}_0$ corresponds to some reference structure.  A pool of
700 individuals is evolved through reproduction and mutation processes
over 500 generations: creation of individuals of the new generation
from individuals of the preceding generation is carried out via
one-point or random cross-over operations while the mutation process,
which re-introduces lost genetic materials and avoids inbreeding, is
realized with a mutation probability of $p_{\rm m} = 5 \%$. For each
state point, 1000 of such independent runs have been carried
out. Finally, the individual with the overall lowest $G$-value, ${\cal
  I}_{\rm min}$, is considered to be the solution of the GA. To
account for the limited accuracy caused by the encoding procedure, the
parameters of ${\cal I}_{\rm min}$ were refined via a final Powell
optimization algorithm \cite{Pow64}.

\subsection{Close-packed structures}
\label{subsec:th_cp}

The limiting case at high pressure is always a lattice where the hard
cores of the particles arrange in a close-packed structure. Thus the
slope of the line that expresses the linear dependence of $g^\star$ on
$P^\star$ in the $(g^\star, P^\star)$-plane is obviously given by
$1/\rho^\star_{\rm max} = 1/\sqrt{2}$.  However, the intercept of this
line, i.e., the energy of this arrangement, $e^\star$, requires more
careful considerations. Below we will give evidence, that the
square-shoulder system is able to self-organize not only in the
well-known close-packed scenarios, i.e., in fcc or hcp lattices, but
also in more complex structures \cite{rareEarth}. We emphasize that
the square-shoulder system serves -- due to its flat plateau and due
to the finite range of the corona -- as an antenna that is able to
identify in a very sensitive way between competing particle
arrangements.

In an effort to find for a given value of $\lambda$ the energetically
most favorable close-packed arrangement of the particles we proceed
as follows. We consider the lattice as being built up by periodically
repeated stacking sequences of $n_l$ hexagonally close-packed layers,
introducing for convenience the conventional labels, $A$, $B$, and $C$
\cite{Ash76}. A stacking sequence of $n_l$ layers can therefore be
described by a string of $n_l$ of these symbols. The trivial
close-packed arrangements, fcc and hcp, are thus characterized by the
sequences $ABC$ (with $n_l = 3$) and $AB$ (with $n_l = 2$).  For a
given value of $n_l$ we consider all possible stacking sequences of
length $n_l$; without loss of generality we start all sequences with
the label $A$. Some of the proposed sequences {\it have} to be ruled
out: this is the case when two neighboring layers carry the same
index. Some of them {\it can} be ruled out: this is, for instance, the
case when symmetry considerations reveal that two different stacking
sequences lead to the same lattice.

Pursuing this strategy we find for the smallest $n_l$-values the
following situation: for $n_l = 2$ we have only the hcp structure
($AB$) and for $n_l = 3$ we recover the fcc lattice ($ABC$). Also for
four- and five-layer stackings only one representative remains: $ABAC$
and $ABABC$ can be identified, respectively. At $n_l = 6$, we
encounter for the first time two non-equivalent stacking sequences,
namely $ABABAC$ and $ABACBC$.  With increasing $n_l$ the number of
possible stacking sequences increases drastically. For instance, for
$\lambda = 4.5$, where we have considered stackings with up to 12
layers, we were able to identify 136 different sequences. A
comprehensive table of possible stacking sequences for a given value
of $n_l$ is presented in \cite{Pau08th}.

Finally, for a given value of $\lambda$ we include a sufficiently
large number of layers and evaluate and compare the energies $e^\star$
of all candidate stackings. The finite range of the shoulder and its
flat energy plateau help to reduce the numerical effort
considerably. On most occasions we encounter degeneracy, i.e., two (or
even more) different stacking sequences are characterized by exactly
the same value of $e^\star$. In such cases, we consider the shortest
among these stacking sequences to be the energetically most favorable
configuration with the only exception that we favor fcc to hcp
\cite{fccFavor}.

\section{Results}
\label{sec:results}

\subsection{Close-packed structures}
\label{subsec:res_cp}

\begin{figure}[!tbh]
      \begin{center}
      \begin{minipage}[t]{8.5cm}
      \includegraphics[width=8.4cm, clip] {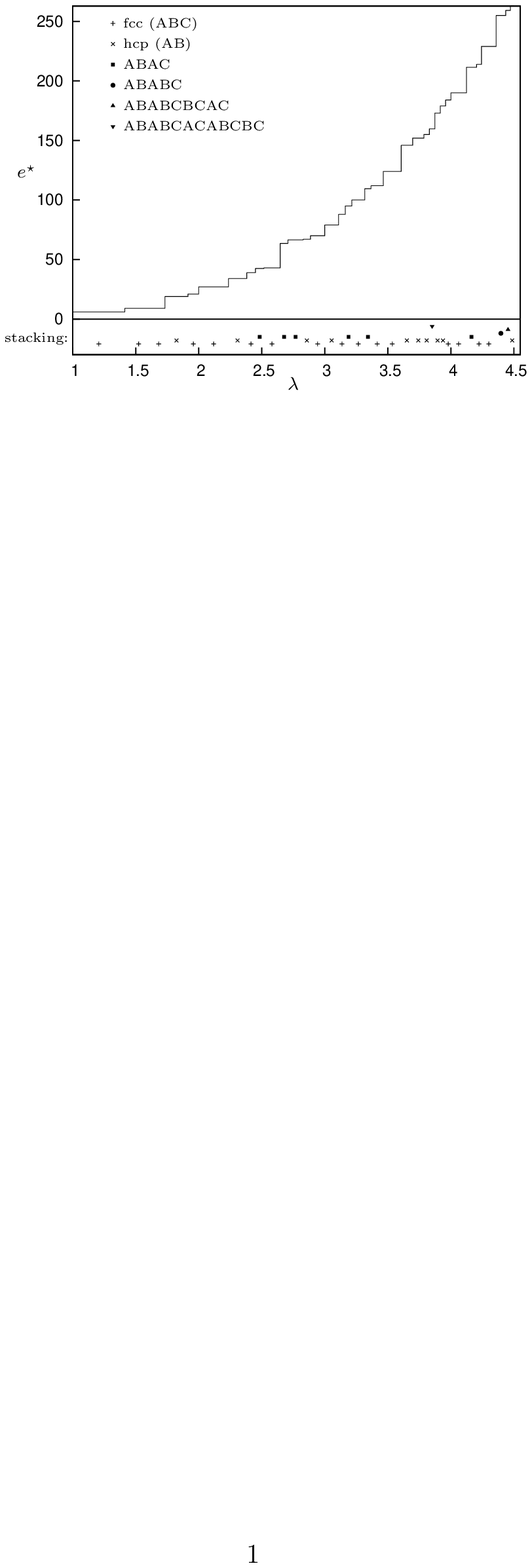}
      \end{minipage}
      \end{center}
\caption{Fig.~\ref{fig:close_pack}: Energy per particle, $e^\star$, for the simplest,
  energetically most favorable close-packed particle arrangements for
  the square-shoulder system as a function of $\lambda$ (full
  curve). Symbols specify the stacking sequences as labeled; see also
  text.}
\label{fig:close_pack}
\end{figure}

With the above considerations in mind we can now identify the
equilibrium close-packed structures for the square-shoulder system as
they occur at high pressure values. These particle arrangements are
summarized in Figure \ref{fig:close_pack}, starting from $\lambda = 1$
(corresponding to hard spheres) and extending to a shoulder width of
$\lambda = 4.5$. The figure contains the energy $e^\star$ of the
respective structures and symbols characterize their stacking
sequences.

As long as $1 < \lambda < \sqrt{2}$, only nearest neighbor
interactions have to be considered. All stackings are characterized by
the same number of overlapping coronae; the energy of a tagged
particles amounts to half the number of nearest neighbors, i.e.,
$e^\star = 6$. Although for $\sqrt{2} < \lambda < 2 \sqrt{2/3}$, the
second nearest neighbors start to play a role, we still obtain for
all possible stackings the same value for $e^\star$, namely $e^\star =
9$.  For $2 \sqrt{2/3} < \lambda < \sqrt{3}$, the different stackings
are characterized by different energy values; among these, the fcc
structure is one of the stackings with the lowest $e^\star$-value, namely
$e^\star = 9$, while all other stackings are energetically equal or
less favorable. Therefore in this $\lambda$-interval, fcc remains the
simplest, energetically most favorable structure. For $\lambda =
\sqrt{3}$, a hcp lattice with $e^\star = 19$ becomes the simplest
structure with the lowest $e$-value. Further stacking sequences can be
extracted from Figure \ref{fig:close_pack} for $\lambda$-values up to
4.5.  This figure shows that for a few $\lambda$-intervals also
stacking sequences other than fcc or hcp are obtained as the
energetically most favorable close-packed particle arrangements at
high pressure.

\subsection{Configurations that minimize the Gibbs free energy}
\label{subsec:res_mecs}

Once we have determined the limiting high pressure MECs, we can
proceed to the identification of the whole sequence of MECs as a
function of the pressure. This is done in the following three
subsections where we have considered square-shoulder systems with a
short ($\lambda = 1.5$), an intermediate ($\lambda = 4.5$), and a
large ($\lambda = 10$) shoulder width. Abbreviations of the underlying
lattices that are used in the text and in the Figures are summarized
in Table \ref{tab:abbrev}.

For $\lambda= 1.5$ we shall give a detailed geometrical interpretation
of these particle arrangements; this will provide clear evidence about
the system's strategy to arrange the particles at given pressure in
such a way as to minimize the number of overlapping shoulders and to
maximize at the same time the density. Although we will not be able to
pursue these geometrical considerations in full detail for the other
$\lambda$-values, we will be able to identify an emerging sequence of
structural archetypes as we increase the pressure: while at low
pressure particles tend to arrange in clusters, which then populate
the positions of regular lattices, we encounter with increasing
pressure columnar, lamellar, and, finally, compact structures. With a
few exceptions this rule is obeyed for $\lambda = 4.5$, while it is
strictly followed for $\lambda = 10$. For a more detailed presentation
we refer to \cite{Pau08th}.

\begin{figure}[!tbh]
      \begin{center}
      \begin{minipage}[t]{8.5cm}
      \includegraphics[width=8.4cm, clip] {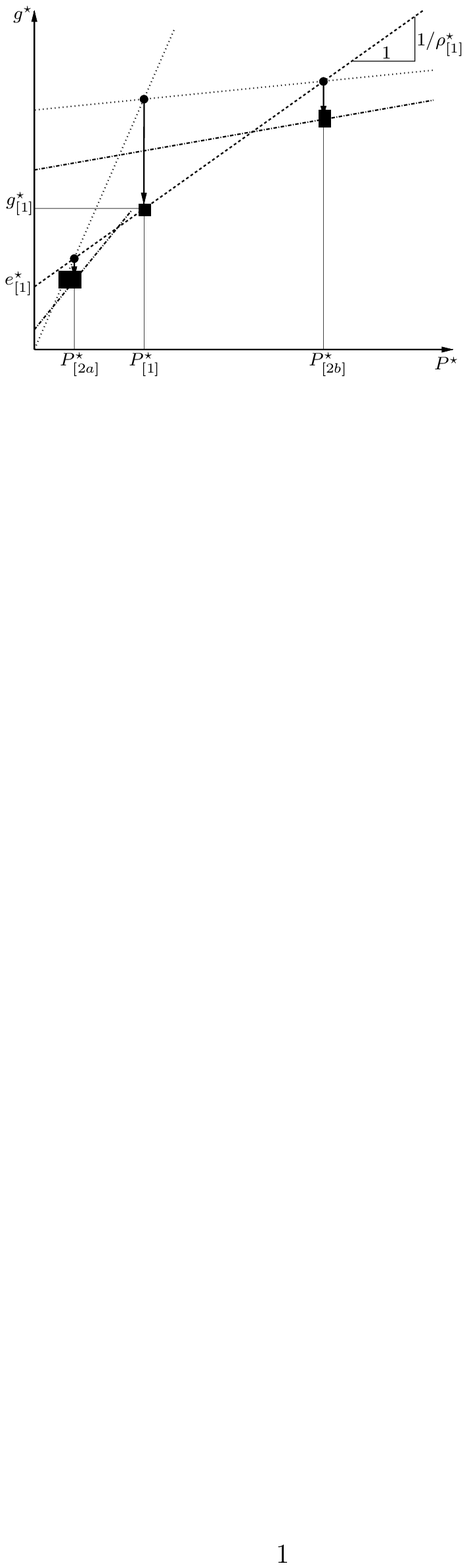}
      \end{minipage}
      \end{center}
\caption{Schematic representation of our search strategy to identify
  MECs in the $(g^\star, P^\star)$-plane. The dotted lines represent
  $g^\star$ as a function of $P^\star$ for the limiting low and high
  pressure configurations: The vertical arrows represent GA-runs that
  identify, starting from an initial guess (dot), an energetically
  more favorable MEC (square). For details see text.}
\label{fig:schematic}
\end{figure}

Before we present and discuss the sequences of MECs in detail we
briefly outline how we can take benefit from the fact that for this
particular system $g^\star$ is a linear function of $P^\star$.  Our
search algorithm is sketched in Figure \ref{fig:schematic}. In a first
step, we determine the intersection point of the two straight lines in
the $(g^\star, P^\star)$-diagram which represent the high and the low
pressure limiting cases; let the corresponding pressure value be
$P^\star_{[1]}$. At this state point we perform a sequence of GA
searches. In each of these runs we consider a different number of
basis atoms, where the maximum number of basis particles depends on
the value of $\lambda$. This optimization step leads to a new particle
configuration which is characterized by a Gibbs free energy
$g^\star_{[1]}$ that is lower than the one of the intersection point,
by a density $\rho^\star_{[1]}$, and by an energy
$e^\star_{[1]}$. Thus this particle arrangement is at given pressure
$P^\star_{[1]}$ the energetically most favorable one. $e^\star_{[1]}$
and $\rho^\star_{[1]}$ define a new line in the $(g^\star,
P^\star)$-plane; we determine the two intersection points of this line
with the two lines representing the limiting configurations leading to
the pressure values $P^\star_{[2a]}$ and $P^\star_{[2b]}$. At these
two state points we launch new GA searches.  This procedure is
repeated until at none of the intersection points of an iteration step
an energetically more favorable particle arrangement can be
identified.  On one side this procedure avoids a rather time-consuming
scan of the pressure range on a finite grid and thus brings along a
considerable reduction of the number of GA steps and, consequently, of
the computational effort; on the other side this strategy avoids the
risk of simply 'forgetting' MECs. Both features become more and more
important with increasing shoulder width, since the distribution of
MECs over the whole pressure range is highly nonlinear, as can be seen
in Figures \ref{fig:enth_1.5}, \ref{fig:enth_4.5},
and \ref{fig:enth_10}.

This systematic search strategy, in combination with the reliability
of GA-based optimizations, make us confident that the sequences of
MECs that we shall present in the following, are complete.

\subsubsection{Short shoulder width ($\lambda = 1.5$)}

\begin{figure}[!tbh]
      \begin{center}
      \begin{minipage}[t]{8.5cm}
      \includegraphics[width=8.4cm, clip] {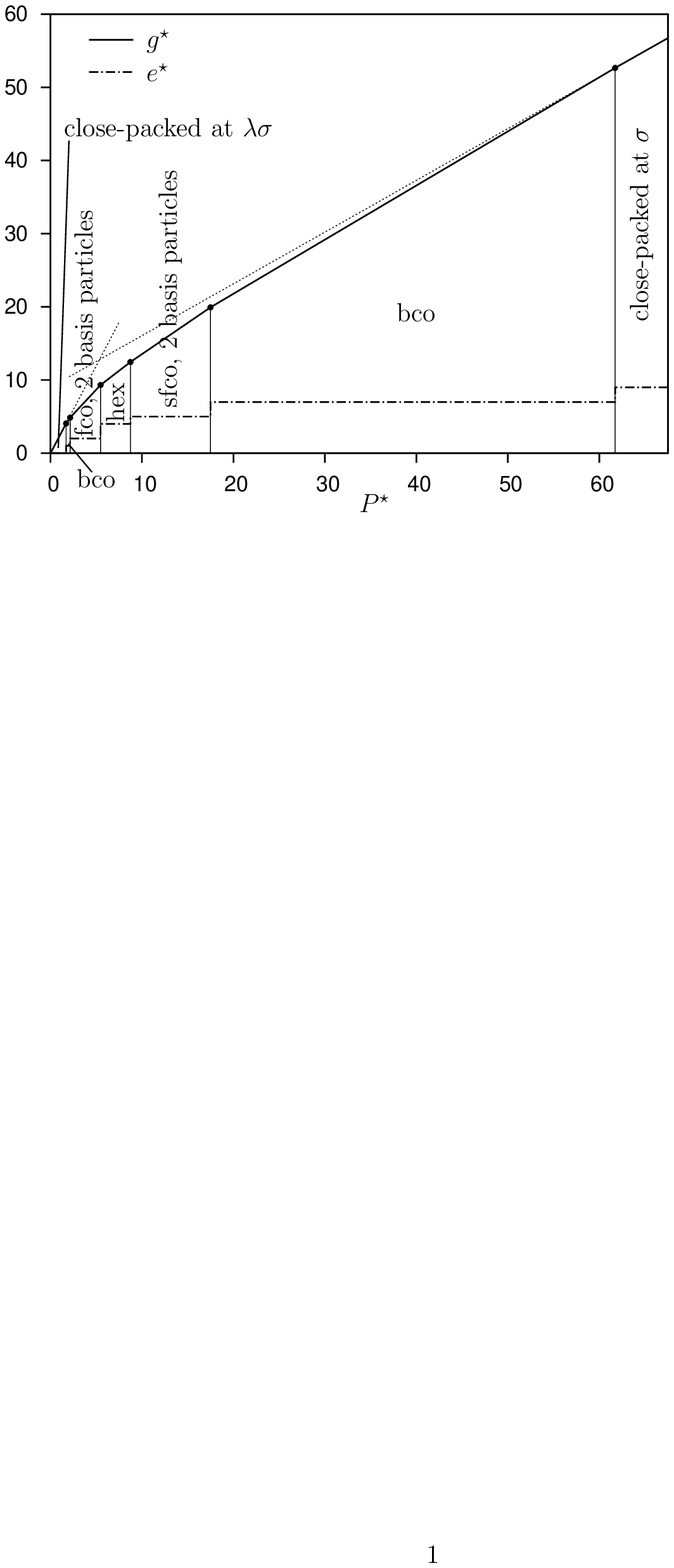}
      \end{minipage}
      \end{center}
\caption{$g^\star$ and $e^\star$ as functions of $P^\star$ for the
  ordered equilibrium structures identified for the square-shoulder
  system with $\lambda = 1.5$, as labeled. The dotted lines indicate
  the low and high pressure limiting configurations (see text). The
  identified lattices are indicated by standard abbreviations (see
  table \ref{tab:abbrev}), including, if required, the number of basis
  particles; see also figure \ref{fig:struct_1.5}.}
\label{fig:enth_1.5}
\end{figure}

\begin{figure}[!tbh]
      \begin{center}
      \begin{minipage}[t]{8.5cm}
      \includegraphics[width=6cm, clip] {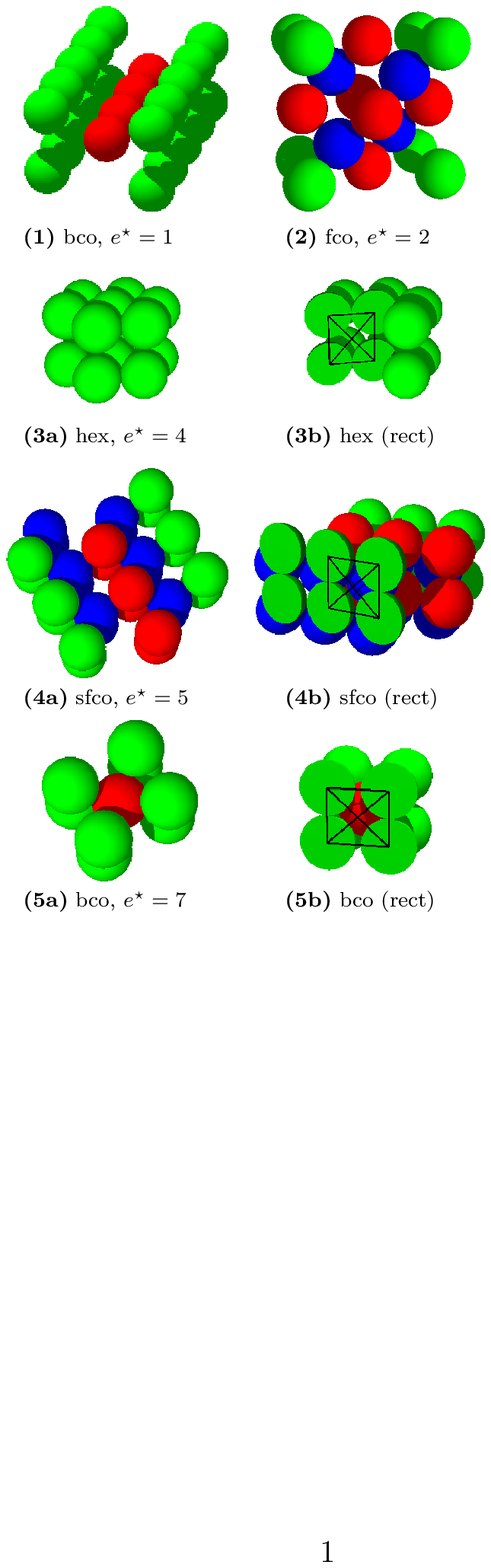}
      \end{minipage}
      \end{center}
\caption{Visualization of the non-trivial ordered equilibrium
  structures for the square-shoulder system with $\lambda =
  1.5$. Structures are characterized by standard abbreviations (see
  table \ref{tab:abbrev}) and their respective $e^\star$-value. Color
  code: green -- particles at the corner positions of the conventional
  unit cell; red -- particles at body- or face-centered positions;
  blue -- additional basis particles.}
\label{fig:struct_1.5}
\end{figure}

The phase diagram (i.e., $g^\star$ and $e^\star$ as functions of
$P^\star$) for the square-shoulder system with $\lambda = 1.5$ is
depicted in Figure \ref{fig:enth_1.5}; the corresponding ordered
equilibrium structures are compiled in Figure \ref{fig:struct_1.5},
except for the trivial low and high pressure structures, where
particles arrange in any close-packed crystal structure, which we take
to be fcc (see discussion above). Further numerical details about the
seven identified MECs are compiled in Table \ref{tab:1.5}. Although we
have considered in our search strategy lattices with up to eight basis
particles, only lattices with at most two basis particles were
identified. The limiting low pressure configuration is characterized
by $e^\star = 0$ and $\rho^\star = \sqrt{2}/\lambda^3 \simeq
0.629$. Further, since $\lambda = 1.5$ is slightly larger than
$\sqrt{2}$, we are above the threshold value (cf. discussion in
subsection \ref{subsec:res_cp}) where the hard cores of the particles
form a close-packed structure and only the coronae of nearest
neighboring particles overlap; thus $e^\star = 9$ and $\rho^\star =
\sqrt{2}$.

As we start our search in the low-pressure regime, the first
non-trivial structure we encounter is a body centered orthorhombic
(bco) structure [Figure \ref{fig:struct_1.5}(1)]. A more detailed
consideration identifies this particle arrangement as a columnar
structure: particles form lanes along which the hard cores are in
direct contact. While, of course, these lanes lead to an
intra-columnar shoulder overlap along the lanes, any other
inter-columnar overlap is avoided; consequently, $e^\star = 1$. Simple
geometric considerations reveal that the edge lengths of the
conventional bco unit cell have the following values: $\sigma$,
$\lambda \sigma$, and $\sqrt{3 \lambda^2 - 1}~\sigma$.

The avoidance of inter-columnar shoulder overlap has to be sacrificed
as the pressure is further increased, leading to a rather compact
structure: we identify a face centered orthorhombic (fco) lattice with
an additional basis particle (color code: blue) [cf. Figure
  \ref{fig:struct_1.5}(2)]. This particle is in direct hard core
contact with its four nearest neighbors: three of them are located at
the faces (color code: red) and one sits at the corner (color code:
green) of the conventional unit cell. Furthermore, this particle is
separated by a distance $\lambda \sigma$ from its second nearest
neighbor (color code: green) which occupies another corner of the
conventional orthorhombic unit cell. Finally, particles at the
smallest faces of the cell (color code: red) are positioned in such a
way that their shoulders touch the shoulders of the particles located
at the corners of the corresponding face of the conventional
orthorhombic unit cell (color code: green). These considerations fix
the edge lengths of the unit cell to be $\sqrt{2 \lambda^2 -
  2}~\sigma$, $\sqrt{2 \lambda^2 + 2}~\sigma$, and $2 \sqrt{4 -
  \lambda^2}~\sigma$.

As we further increase the pressure the particles arrange in a MEC
that can be identified as a lamellar structure, cf. Figure
\ref{fig:struct_1.5}(3a). It can be described as a stacking of
hexagonally close-packed layers which are placed exactly on top of
each other. The nearest neighbor distance is obviously $\sigma$.  The
second nearest neighbors are separated by the inter-layer distance
which is fixed by the requirement that the corona of a tagged particle
touches the shoulders of its twelve third nearest neighbors, located
in the adjacent layers. Thus, this distance amounts to
$\sqrt{\lambda^2 - 1}~\sigma \sim 1.12 \sigma$. It should be pointed
out that particles in nearest and second nearest neighbor distance
form a rectangular particle arrangement [emphasized in Figure
  \ref{fig:struct_1.5}(3b)] that will also be encountered in the
subsequent MECs: if we consider within a layer two particles in close
contact, then they form with the corresponding particles of one of the
adjacent layers a rectangle with edge lengths $\sigma$ and
$\sqrt{\lambda^2 - 1}~\sigma$.

For even higher pressure values only compact structures are
identified. The next MEC can be described as a single face centered
orthorhombic (sfco) lattice with two basis particles, visualized in
Figure \ref{fig:struct_1.5}(4a). The orthorhombic unit cell is built
up by two side faces that have exactly the aforementioned rectangular
shape [formed by particles in green, emphasized in Figure
\ref{fig:struct_1.5}(4b)], while the two larger side faces are each
decorated in their center by an additional particle (color code:
red). Finally, the additional basis particles (color code: blue) are
located in such a way, that they are in direct contact both with the
four particles forming the side faces as well as with the two red
particles located in the other side faces.  Simple geometric
considerations lead to the edge lengths of the orthorhombic cell,
namely: $\sigma$, $\sqrt{\lambda^2 - 1}~\sigma$, and $(\sqrt{3} +
\sqrt{4 - \lambda^2})~\sigma$.

The last non-trivial compact structure is a body centered orthorhombic
lattice [emphasized in Figure \ref{fig:struct_1.5}(5a)]. Again, we can
easily identify the side faces of the conventional unit cell as the
above mentioned rectangular structure [see Figure
  \ref{fig:struct_1.5}(5b)]. In addition, the central particle is in
direct hard core contact with the particles forming the unit
cell. Based on these geometrical considerations the edge lengths of
the unit cell can easily be identified to be $\sigma$,
$\sqrt{\lambda^2 - 1}~\sigma$, and $\sqrt{4 - \lambda^2}~\sigma$.

\subsubsection{Intermediate shoulder width ($\lambda = 4.5$)}

A much larger diversity in the ordered equilibrium structures could be
identified for an intermediate shoulder width of $\lambda = 4.5$. The
limiting low pressure MEC is of course again an fcc structure with a
nearest neighbor distance $\lambda\sigma$ and, hence, $e^\star =
0$. On the other hand, the high-density limiting particle
configuration is an hcp lattice with $e^\star = 263$ (see Figure
\ref{fig:close_pack} and discussion in subsection
\ref{subsec:res_cp}). With the help of the GA we have obtained in
total 33 different MECs over the entire pressure regime. In our
investigations unit cells with up to 10 basis particles have been
considered, in the end only configurations with up to 8 basis
particles were part of the MECs.

A first look at these MECs gives evidence that within this sequence of
MECs we can easily identify the aforementioned four structural
archetypes: at low pressure, the system prefers to form cluster
structures; with increasing pressure, columnar structures are formed,
which then transform into lamellar particle arrangements; finally, at
high pressure, we observe compact structures. This rule, which is
disobeyed only twice for $\lambda = 4.5$, can nicely be understood via
a detailed analysis of the particle arrangements, reflecting the
system's strategy to reduce at a given pressure the number of
overlapping coronae as much as possible (i.e., minimizing $e^\star$)
while maximizing at the same time the particle density.

\begin{figure}[!tbh]
      \begin{center}
      \begin{minipage}[t]{8.5cm}
      \includegraphics[width=8.4cm, clip] {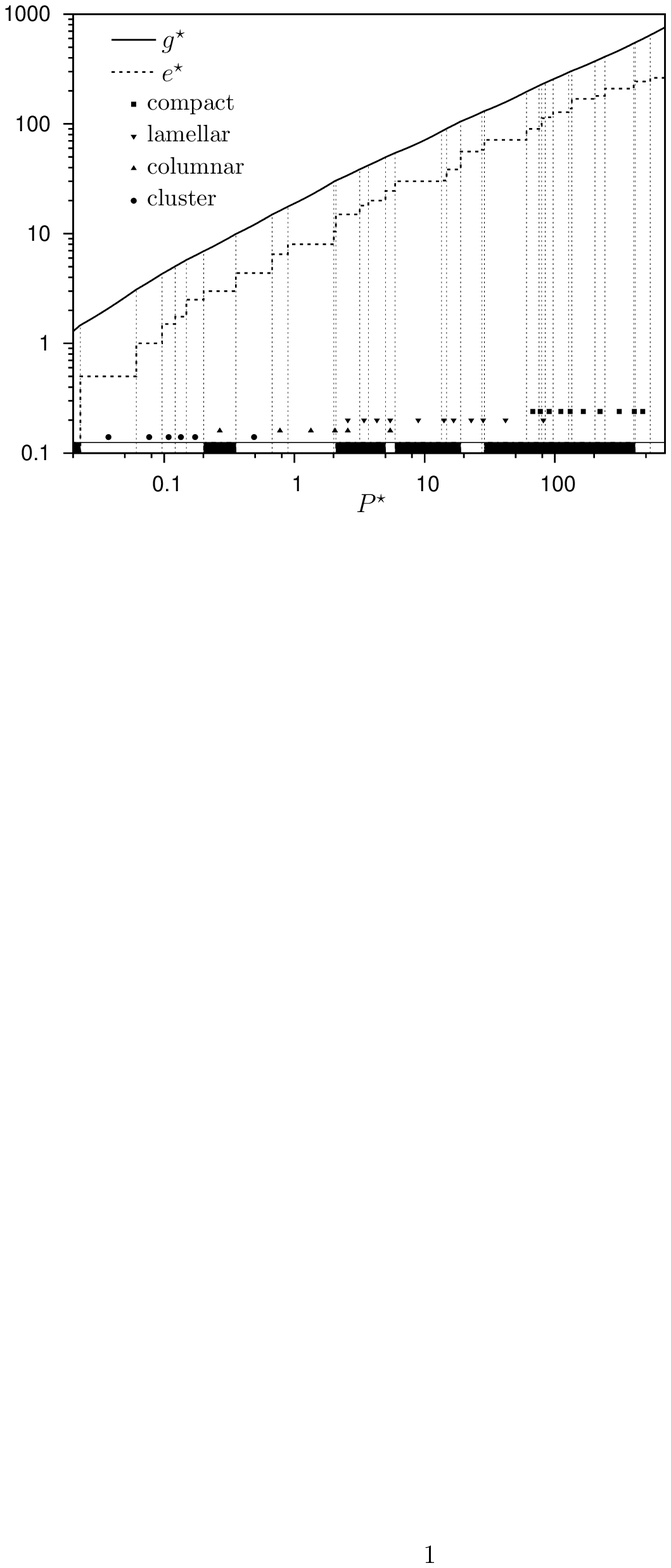}
      \end{minipage}
      \end{center}
\caption{$g^\star$ and $e^\star$ as functions of $P^\star$ on a double
  logarithmic scale for the ordered equilibrium structure identified
  for the square-shoulder system with $\lambda = 4.5$, as labeled;
  note that due to the non-linear scale, the linear dependence between
  $g^\star$ and $P^\star$ is no longer visible. The structural
  archetypes to which a given MEC belongs (cluster, columnar,
  lamellar, or compact structure) are specified by a symbol (as
  labeled). The black horizontal bar at the bottom of the figure
  indicates those ordered particle arrangements where no direct
  contact between the cores of the particles is observed.}
\label{fig:enth_4.5}
\end{figure}

Numerical details about these ordered structures are summarized in
Table \ref{tab:4.5_array}. The phase diagram for the square-shoulder
system with $\lambda = 4.5$ is depicted in Figure
\ref{fig:enth_4.5}. It also contains information to which class of the
four archetypes a particular MEC belongs. Finally, the horizontal bar
at the bottom of the figure indicates those MECs where no direct
contact between the cores of the particles occurs.

\begin{figure*}[!tbh]
  \includegraphics[width=\textwidth,clip] {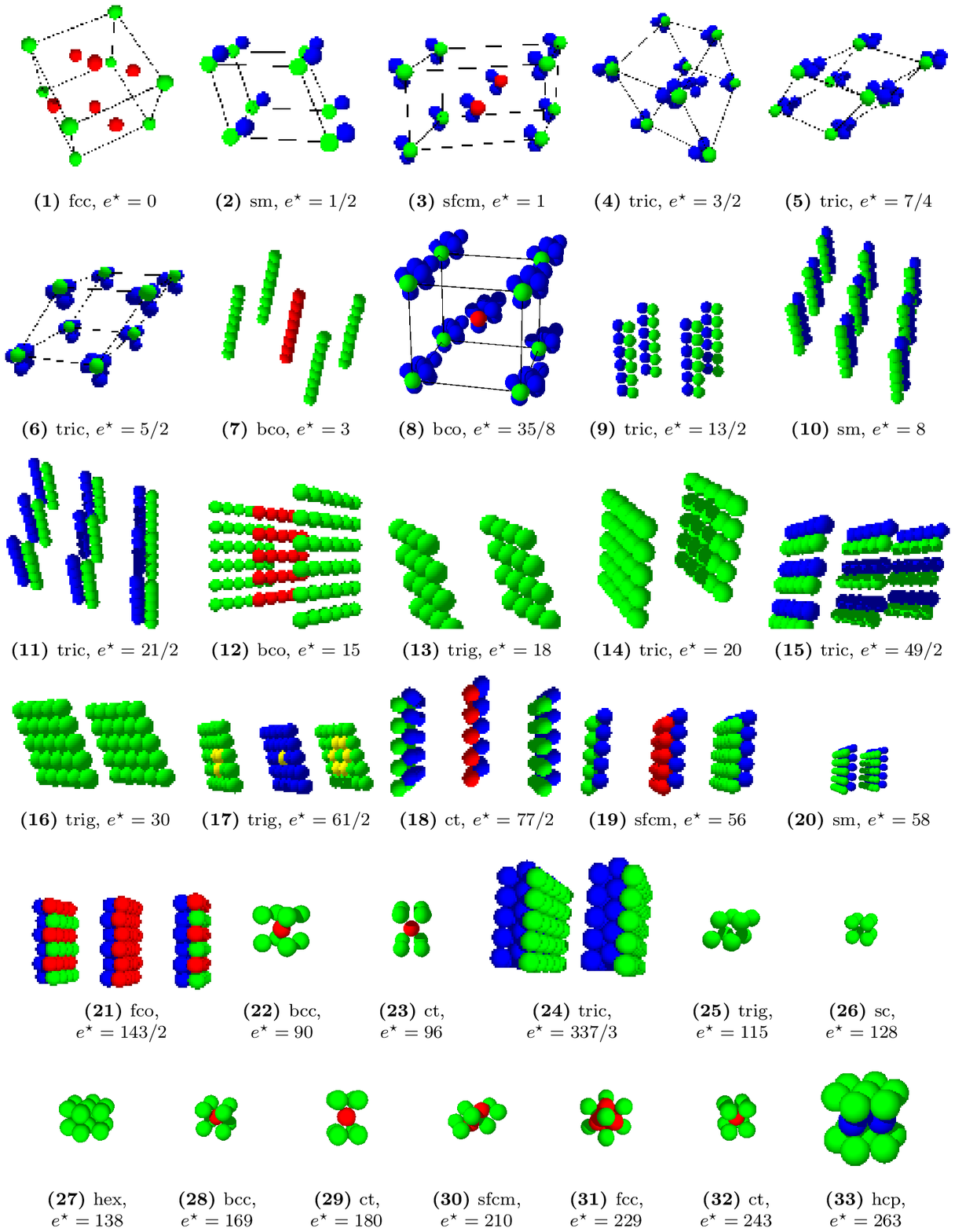}
  \caption{Visualization of all 33 ordered equilibrium structures for
    the square-shoulder system with $\lambda = 4.5$. Structures are
    characterized by standard abbreviations (see table
    \ref{tab:abbrev}) and their respective $e^\star$-value. Color
    code: green -- particles at the corner positions of the
    conventional unit cell; red -- particles at body- or face-centered
    positions; blue -- additional basis particles. The shoulders of
    the yellow particles in panel (17) touch the ones of the other
    yellow particles, located in the neighboring layers.}
  \label{fig:struct_4.5}
\end{figure*}

At low pressure-values (i.e., up to $P^\star \simeq 1.70$) particles
prefer to arrange in ordered clusters of up to eight particles, which,
in turn, populate the positions of crystal lattices. A closer analysis
of these structures reveals a strong interplay between the shape of
the clusters and of the symmetry of the unit cell: the more aspherical
the clusters are the lower is the symmetry of the lattice. This
tendency reflects the system's efforts to avoid to a highest possible
degree a shoulder overlap of neighboring clusters. A nice
visualization of this strategy can, for instance, be observed in the
structure depicted in Figure \ref{fig:struct_4.5}(5): the rather
elongated four-particle clusters are located on a low-symmetry
triclinic lattice; on the other hand, the nearly spherically shaped
eight-particle clusters of the structure depicted in Figure
\ref{fig:struct_4.5}(8) populate the lattice positions of a bco
structure, that has a considerably higher symmetry. A systematic,
quantitative analysis of all cluster structures reveals that for most
of these MECs only rarely shoulder overlaps of neighboring clusters
are observed. A nice example that demonstrates the complexity of
cluster structures is depicted in Figure \ref{fig:struct_4.5}(4). In
this MEC we can identify two different sorts of clusters: tetrahedral
clusters occupy the corners of a triclinic lattice, while the other
four-particle cluster species populates a central position in 
the body of the triclinic cell.

As the pressure is further increased there is a drastic change in the
system's strategy to arrange particles, namely the formation of
columnar structures, where particles self-organize in lanes. This
leads to a considerable energetic penalty, since -- due to the short
inter-particle distance within the columns -- an appreciable number of
overlapping shoulders is induced; at the same time a rather high
particle density is guaranteed along these lanes which, in turn,
contributes to e reduction in the Gibbs free energy. Simultaneously,
the system tries to compensate for this high energetic cost within the
lanes by arranging these columns in such a way as to minimize the
inter-columnar shoulder overlap. This strategy leads first to the
formation of single-columnar structures [as depicted in Figure
  \ref{fig:struct_4.5}(7)], and later, as the pressure is increased,
to double-columnar particle arrangements [cf. Figure
  \ref{fig:struct_4.5}(9--11)]. We point out that within the lanes
particles are only in direct contact at sufficiently high pressure; in
double-columnar structures, particles of adjacent columns are always
in direct contact. The system's strategy to avoid shoulder overlap
between the lanes can nicely be traced in a closer analysis: for the
structures depicted in Figures \ref{fig:struct_4.5}(7,9,10), no corona
overlap between the single or double strands is observed; only at
sufficiently high pressure -- cf. Figure \ref{fig:struct_4.5}(11) --
the coronae of different neighboring double columns start to overlap.

For pressures values above $P^\star \gtrsim 3$, the system has to
search for new ideas of how to minimize the Gibbs free energy. Now the
change to a new strategy is considerably smoother than the preceding
one: in an effort to cope with the increasing pressure the system
forms lamellar structures -- cf. Figures \ref{fig:struct_4.5}(12--21)
and Figure \ref{fig:struct_4.5}(24). These MECs emerge from columnar
structures as the columns approach each other, forming thereby
lamellae; some intermediate stages of this transition can be observed
in Figure \ref{fig:struct_4.5}(12) and Figure
\ref{fig:struct_4.5}(15). Within the lamellar structure the system's
strategy is obvious. First optimize the packing inside a layer,
leading to hexagonal particle arrangements inside a lamella: while at
low pressure values [Figures \ref{fig:struct_4.5}(13,14)] particles
are more loosely packed, they are forced to form a nearly hexagonally
close-packed structure with a nearest neighbor distance of $\simeq
1.03\sigma$ at higher pressure [Figures
  \ref{fig:struct_4.5}(16,17)]. Particular attention should be
dedicated to the latter structure: the three neighboring, parallel
planes depicted in Figure \ref{fig:struct_4.5}(17) are {\it not}
equally spaced; the two different emerging inter-lamellar distances
are rather governed by the fact that the shoulders of the particles
marked in yellow located in the three neighboring layers touch.  If
the possibility for optimizing the packing within a single lamella has
been exhausted, the system starts to form double layers [cf. Figures
  \ref{fig:struct_4.5}(18--21)] or even triple layers [cf. Figure
  \ref{fig:struct_4.5}(24)]. A closer analysis of the double-layer
structure reveals a very complex strategy which we try to explain as
follows. We consider two pairs of neighboring double-layers. On one
hand we observe shoulder overlap between single layers (belonging to
different pairs) that face each other: for instance, in Figure
\ref{fig:struct_4.5}(19), the layer formed by blue particles in the
left-most layer pair and the layer formed by red particles in the
central layer pair; on the other hand the distance between pairs of
layers is chosen in such a way as to avoid shoulder overlap of single
layers that do not face each other: for instance, in Figure
\ref{fig:struct_4.5}(19), the layers formed by green particles and the
layers formed by red particles.  These observations turn out to be
valid for all double layer structures that have been identified for
this particular shoulder width.

Finally, we enter for high pressure values the regime of compact
structures, characterized, in general, by a large number of nearest
neighbors. In most of these MECs direct core contact is avoided
(cf. horizontal bar in Figure \ref{fig:enth_4.5}), only in the
high-pressure regime, where a ct-lattice [cf.~Figure
  \ref{fig:struct_4.5}(32)] and, finally, an hcp-structure [the
  limiting case for $\lambda = 4.5$, see Figure
  \ref{fig:struct_4.5}(33)] are the respective MECs, the cores are in
direct contact. Again, with simple geometric considerations, the
system's strategy to form MECs can be traced back to avoiding
unnecessary shoulder overlap while maximizing the particle density.

\subsubsection{Large shoulder width ($\lambda = 10$)}\label{subsec:10}

Finally, we consider the case of a large shoulder width for which we
have chosen $\lambda = 10$. Now the hard core region is relatively
small with respect to the shoulder range. Thus at low densities the
core plays a minor role and the system becomes closely related to the
penetrable sphere model (PSM) \cite{Lik98Wat}. The PSM belongs to a
class of soft matter systems where particles are able to solidify in
so called cluster phases \cite{Lik01Lan}, i.e., where stable clusters
of particles form, which, in turn, populate the positions of periodic
lattices. Evidence for this particular phase behavior has been given
in density functional based investigations and in computer simulations
for the PSM \cite{Fal08} and via purely theoretical considerations,
combined with computer simulations for a closely related model
potential \cite{Mla06PRL,Mla06PRLERR,Lik07,Mla07Got}. As we will show
below, such cluster phases can also be observed for the
square-shoulder system at low pressure values where the hard cores of
the particles still have a negligible effect on the properties of the
system.

\begin{figure}[!tbh]
      \begin{center}
      \begin{minipage}[t]{8.5cm}
      \includegraphics[width=8.4cm, clip] {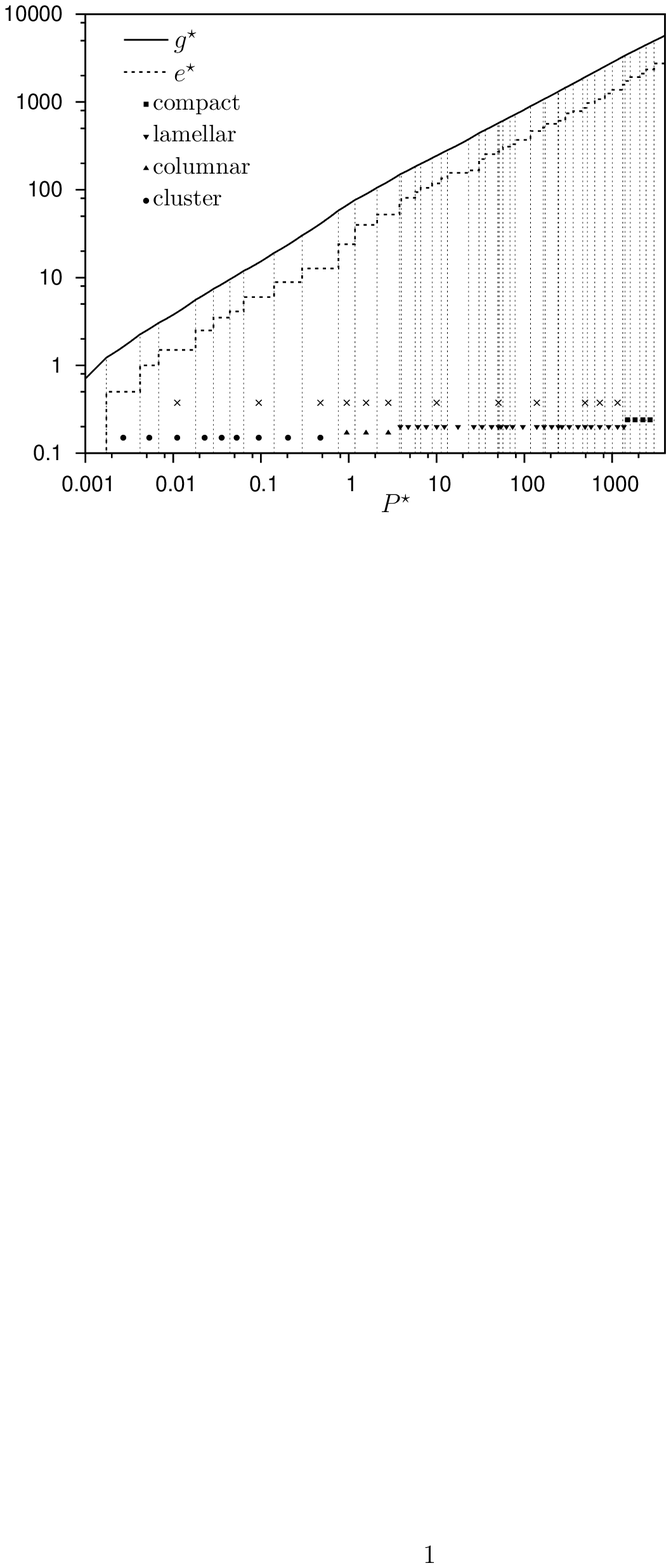}
      \end{minipage}
      \end{center}
\caption{$g^\star$ and $e^\star$ as functions of $P^\star$ on a double
  logarithmic scale for the ordered equilibrium structures identified
  for the square-shoulder system with $\lambda = 10$, as labeled; note
  that due to the non-linear scale, the linear dependence between
  $g^\star$ and $P^\star$ is no longer visible. The structural
  archetypes to which a given MEC belongs (cluster, columnar,
  lamellar, or compact structure) are specified by a symbol (as
  labeled). Ordered equilibrium structures marked by crosses are
  visualized in Figure \ref{fig:struct_10}.}
\label{fig:enth_10}
\end{figure}

Since the MECs are expected to be rather complex we have considered up
to 29 basis particles in our GA based search strategy, up to 22
appeared in the MECs. In total we have identified as much as 47 MECs,
i.e., a relatively large number which makes a detailed discussion and
interpretation of the structures impossible. The thermodynamic
properties of all these MECs are displayed in Figure
\ref{fig:enth_10}. We point out that the high pressure limiting
configuration is a hcp lattice with $e^\star = 2947$. For the case
$\lambda = 10$ the rule for the sequence of structural archetypes
(cluster -- columnar -- lamellar - compact structures) is strictly
obeyed (see symbols in Figure \ref{fig:enth_10}).

\begin{figure*}[!tbh]
  \includegraphics[width=\textwidth,clip] {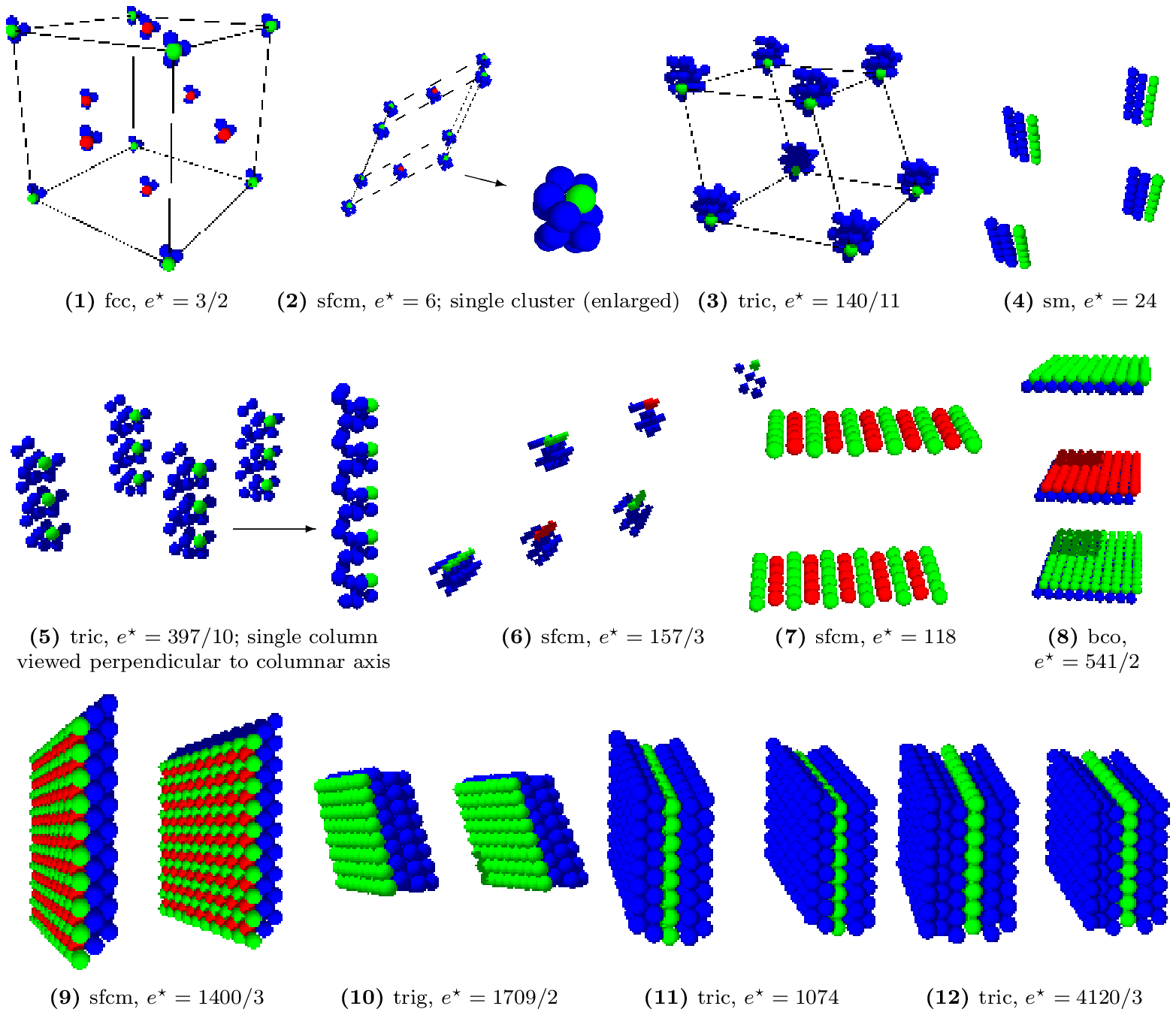}
  \caption{Visualization of a selection of the 47 ordered equilibrium
    structures for the square-shoulder system with $\lambda =
    10$. Structures are characterized by standard abbreviations (see
    table \ref{tab:abbrev}) and their respective
    $e^\star$-value. Color code: green -- particles at the corner
    positions of the conventional unit cell; red -- particles at body-
    or face-centered positions; blue -- additional basis particles.}
  \label{fig:struct_10}
\end{figure*}

As expected cluster structures emerge at low pressure values. A few
examples of the ten cluster structures that have been identified are
depicted in Figures \ref{fig:struct_10}(1--3). The clusters can
contain as many as 22 particles [e.g., in the structure in depicted in
Figure \ref{fig:struct_10}(3)] and are arranged in complex structures.
An example for a typical cluster is depicted in Figure
\ref{fig:struct_10}(2); in general the intra-cluster arrangement of
the particles turns out to be irregular.

At $P^\star \simeq 0.76$ the transition to the columnar structures
occurs. The relatively large shoulder width allows for a large variety
of columnar morphologies, including multi-columnar arrangements or
complex helical columns -- cf. Figures \ref{fig:struct_10}(4--6).  In
Figure \ref{fig:struct_10}(5) a side view of a single column gives
evidence of its complex internal structure. Ten basis particles were
required to parameterize this MEC; a single column can be considered
to be built up by a sequence of aligned clusters. We point out that
also in experiment helical columns were have been observed for a
particular class of colloidal particles \cite{Cam05}. Examples for
multi-columnar arrangements are the triple-columns displayed in Figure
\ref{fig:struct_10}(4), or the MEC shown in Figure
\ref{fig:struct_10}(6): here six parallel single-columns that are
nearly in close contact are aligned in parallel to build the
sixfold-column, as can be seen from the right-most column, where the
direction of projection has been chosen to be parallel to the columnar
axis.

Most of the MECs identified for the case $\lambda = 10$ have lamellar
character: in total we have identified as much as 28 lamellar
MECs. Again, we observe a similar strategy as the one identified for
$\lambda = 4.5$: first the particle arrangement within the single
layer structures is optimized; then, if this possibility for
close-packed arrangements is exhausted, multi-layer structures are
formed. The large shoulder width is responsible both for the large
inter-layer distance as well as the close contact within groups of
lamellae: One has the impression that the large range of the shoulder
compactifies adjacent layers, bringing them in direct contact, while
maximizing at the same time the distance between these groups of
layers [cf. Figures \ref{fig:struct_10}(8--12)].

Finally, we enter the regime of compact structures. Since they
resemble very closely those MECs that have been identified for
$\lambda = 4.5$, we do not present them here.

\section{Conclusions}
\label{sec:conclu}

In this contribution we have thoroughly investigated the phase diagram
of the square-shoulder system at $T = 0$, taking into account a short,
an intermediate, and a large shoulder width. Measuring the range of
the corona in terms of $\lambda \sigma$ (where $\sigma$ is the hard
core diameter) we have assumed the following specific values for the
three cases: $\lambda = 1.5$, $\lambda = 4.5$, and $\lambda =
10$. Investigating the system in the NPT ensemble we have searched for
ordered particle configurations that minimize the Gibbs free energy;
this means that the internal energy is minimized while the particle
density is simultaneously maximized. These particle arrangements have
been identified by means of a search strategy that is based on ideas
of genetic algorithms. With this reliable, flexible, and efficient
optimization tool at hand and taking benefit of the simple functional
form of the inter-particle potential, which considerably facilitates
both numerical calculations as well as geometrical interpretations, we
give evidence that the sequences of emerging particle configurations
of minimum energy are complete.

A first look on the total of configurations gives clear indications
that the formation of the particle arrangements follows well-defined
rules as the pressure is increased: while at low pressure values the
system prefers to form clusters, which, in turn, populate the
positions of low-symmetry lattices, we encounter at medium pressure
values columnar and then lamellar structures. As a rule of the thumb
we found, the distances between clusters, columns and lamellae are
always roughly equal to the shoulder width. Finally, at high pressure
values, rather compact particle configurations are identified, that
are in general characterized by a large number of nearest
neighbors. While this rule for the structural ordering might be still
less obvious for small values of $\lambda$, it becomes more apparent
with increasing range of the corona: at the intermediate
$\lambda$-value of $4.5$, it is disobeyed only at two occasions, and
for $\lambda = 10$ the ordered particle configurations fully match
this rule.

The large variety of ordered equilibrium configurations that have been
identified is overwhelming. It represents an impressive example of the
capacity and propensity of soft matter particles to self-organize in
highly non-trivial structures. For demonstration we pick out two
particular examples: first a cluster structure, where the clusters are
composed of 22 particles, which, in turn, populate the positions of a
triclinic lattice; second a columnar structure, where the particles
align in a complex, helical column.

The particular shape of the repulsive shoulder, i.e., its flat
energetic plateau in combination with the well-defined range of the
corona, provides in addition the unique possibility to {\it
  understand} the system's strategy to form these complex
structures. A detailed structural and energetical analysis of the
emerging configurations reveals that the shoulder width plays the
dominant role in this process; however, also the hard core can have
considerable influence on the structure formation since it represents
a lower boundary to inter-particle distances. Thus, this study can be
viewed as a pedagogical example which provides a deeper insight why
particles arrange at a given state point in a particular structure, a
knowledge, which might be helpful in the investigations of
self-assembly processes of systems with more complex inter-particle
potentials.

What are the next steps? The most obvious extension should be directed
towards the investigations of the phase diagram at finite temperature,
i.e., one should address the question which of these configurations
will 'survive' at $T > 0$. The theoretical route to this answer is
quite straightforward: one has to 'simply' merge the proposed search
strategy with a suitable method to evaluate the thermodynamic
properties of the system at a finite temperature. The obvious
candidate to evaluate the thermodynamic properties of the system is of
course classical density functional theory \cite{Eva92}. However, we
have to raise immediately two serious concerns. First, no reliable
density functional format for the square-shoulder system is available
at present; treating, alternatively, the hard core within a suitable
fundamental measure theory format (as, e.g., the one proposed in
\cite{Rot02}) and considering the shoulder by a mean-field type
perturbative approach, risks to provide data that are not sufficiently
accurate to reliably distinguish between energetically competitive
structures. Second, in search strategies based on ideas of genetic
algorithms, the evaluation of the fitness function for an individual
represents the numerical bottleneck in this approach. According to our
experience, a combination of classical density functional theory with
this particular search strategy might still be too time-consuming for
present-day computers.  Alternatively to the theoretical route one
might of course perform computer simulations and we point out that
recently investigations in this direction have been performed on a
related system \cite{Fom06}.

\section*{Acknowledgements}

The authors are indebted to Julia Fornleitner and Dieter Gottwald
(both Wien) for stimulating discussions and for computational
aid. Financial support by the Austrian Science Foundation under
Proj. Nos. W004, P17823-N08, and P19890-N16 is gratefully
acknowledged.

\eject

\begin{thebibliography}{99}
\expandafter\ifx\csname natexlab\endcsname\relax\def\natexlab#1{#1}\fi
\expandafter\ifx\csname bibnamefont\endcsname\relax
  \def\bibnamefont#1{#1}\fi
\expandafter\ifx\csname bibfnamefont\endcsname\relax
  \def\bibfnamefont#1{#1}\fi
\expandafter\ifx\csname citenamefont\endcsname\relax
  \def\citenamefont#1{#1}\fi
\expandafter\ifx\csname url\endcsname\relax
  \def\url#1{\texttt{#1}}\fi
\expandafter\ifx\csname urlprefix\endcsname\relax\def\urlprefix{URL }\fi
\providecommand{\bibinfo}[2]{#2}
\providecommand{\eprint}[2][]{\url{#2}}

\bibitem[{\citenamefont{Hemmer and Stell}(1970)}]{Hem70}
\bibinfo{author}{\bibfnamefont{P.~C.} \bibnamefont{Hemmer}} \bibnamefont{and}
  \bibinfo{author}{\bibfnamefont{G.}~\bibnamefont{Stell}},
  \bibinfo{journal}{Phys. Rev. Lett.} \textbf{\bibinfo{volume}{24}},
  \bibinfo{pages}{1284} (\bibinfo{year}{1970}).

\bibitem[{\citenamefont{Kincaid
  et~al.}(1976{\natexlab{a}})\citenamefont{Kincaid, Stell, and Hall}}]{Kin76a}
\bibinfo{author}{\bibfnamefont{J.~M.} \bibnamefont{Kincaid}},
  \bibinfo{author}{\bibfnamefont{G.}~\bibnamefont{Stell}}, \bibnamefont{and}
  \bibinfo{author}{\bibfnamefont{C.~K.} \bibnamefont{Hall}},
  \bibinfo{journal}{J. Chem. Phys.} \textbf{\bibinfo{volume}{65}},
  \bibinfo{pages}{2161} (\bibinfo{year}{1976}{\natexlab{a}}).

\bibitem[{\citenamefont{Kincaid
  et~al.}(1976{\natexlab{b}})\citenamefont{Kincaid, Stell, and
  Goldmark}}]{Kin76b}
\bibinfo{author}{\bibfnamefont{J.~M.} \bibnamefont{Kincaid}},
  \bibinfo{author}{\bibfnamefont{G.}~\bibnamefont{Stell}}, \bibnamefont{and}
  \bibinfo{author}{\bibfnamefont{E.}~\bibnamefont{Goldmark}},
  \bibinfo{journal}{J. Chem. Phys.} \textbf{\bibinfo{volume}{65}},
  \bibinfo{pages}{2172} (\bibinfo{year}{1976}{\natexlab{b}}).

\bibitem[{\citenamefont{Rasc{\'o}n et~al.}(1997)\citenamefont{Rasc{\'o}n,
  Velasco, Mederos, and Navascu{\'e}s}}]{Ras97}
\bibinfo{author}{\bibfnamefont{C.}~\bibnamefont{Rasc{\'o}n}},
  \bibinfo{author}{\bibfnamefont{E.}~\bibnamefont{Velasco}},
  \bibinfo{author}{\bibfnamefont{L.}~\bibnamefont{Mederos}}, \bibnamefont{and}
  \bibinfo{author}{\bibfnamefont{G.}~\bibnamefont{Navascu{\'e}s}},
  \bibinfo{journal}{J. Chem. Phys.} \textbf{\bibinfo{volume}{106}},
  \bibinfo{pages}{6689} (\bibinfo{year}{1997}).

\bibitem[{\citenamefont{Bolhuis and Frenkel}(1997)}]{Bol97}
\bibinfo{author}{\bibfnamefont{P.}~\bibnamefont{Bolhuis}} \bibnamefont{and}
  \bibinfo{author}{\bibfnamefont{D.}~\bibnamefont{Frenkel}},
  \bibinfo{journal}{J. Phys.: Condens. Matter} \textbf{\bibinfo{volume}{9}},
  \bibinfo{pages}{381} (\bibinfo{year}{1997}).

\bibitem[{\citenamefont{Lang et~al.}(2000)\citenamefont{Lang, \mbox{C. N.}
  Likos, Watzlawek, and L{\"o}wen}}]{Lang00}
\bibinfo{author}{\bibfnamefont{A.}~\bibnamefont{Lang}},
  \bibinfo{author}{\bibnamefont{\mbox{C. N.} Likos}},
  \bibinfo{author}{\bibfnamefont{M.}~\bibnamefont{Watzlawek}},
  \bibnamefont{and}
  \bibinfo{author}{\bibfnamefont{H.}~\bibnamefont{L{\"o}wen}},
  \bibinfo{journal}{J. Phys.: Condens. Matter} \textbf{\bibinfo{volume}{12}},
  \bibinfo{pages}{5087} (\bibinfo{year}{2000}).

\bibitem[{\citenamefont{Velasco et~al.}(2000)\citenamefont{Velasco, Mederos,
  Navascu{\'e}s, Hemmer, and Stell}}]{Vel00}
\bibinfo{author}{\bibfnamefont{E.}~\bibnamefont{Velasco}},
  \bibinfo{author}{\bibfnamefont{L.}~\bibnamefont{Mederos}},
  \bibinfo{author}{\bibfnamefont{G.}~\bibnamefont{Navascu{\'e}s}},
  \bibinfo{author}{\bibfnamefont{P.~C.} \bibnamefont{Hemmer}},
  \bibnamefont{and} \bibinfo{author}{\bibfnamefont{G.}~\bibnamefont{Stell}},
  \bibinfo{journal}{Phys. Rev. Lett.} \textbf{\bibinfo{volume}{85}},
  \bibinfo{pages}{122} (\bibinfo{year}{2000}).

\bibitem[{\citenamefont{Jagla}(1998)}]{Jag98}
\bibinfo{author}{\bibfnamefont{E.}~\bibnamefont{Jagla}},
  \bibinfo{journal}{Phys. Rev. E} \textbf{\bibinfo{volume}{58}},
  \bibinfo{pages}{1478} (\bibinfo{year}{1998}).

\bibitem[{\citenamefont{Jagla}(1999{\natexlab{a}})}]{Jag99JCP}
\bibinfo{author}{\bibfnamefont{E.}~\bibnamefont{Jagla}}, \bibinfo{journal}{J.
  Chem. Phys.} \textbf{\bibinfo{volume}{110}}, \bibinfo{pages}{451}
  (\bibinfo{year}{1999}{\natexlab{a}}).

\bibitem[{\citenamefont{Sadr-Lahijany et~al.}(1998)\citenamefont{Sadr-Lahijany,
  Scala, Buldyrev, and Stanley}}]{Sad98}
\bibinfo{author}{\bibfnamefont{M.~R.} \bibnamefont{Sadr-Lahijany}},
  \bibinfo{author}{\bibfnamefont{A.}~\bibnamefont{Scala}},
  \bibinfo{author}{\bibfnamefont{S.~V.} \bibnamefont{Buldyrev}},
  \bibnamefont{and} \bibinfo{author}{\bibfnamefont{H.~E.}
  \bibnamefont{Stanley}}, \bibinfo{journal}{Phys. Rev. Lett.}
  \textbf{\bibinfo{volume}{81}}, \bibinfo{pages}{4895} (\bibinfo{year}{1998}).

\bibitem[{\citenamefont{Jagla}(1999{\natexlab{b}})}]{Jag99JCPb}
\bibinfo{author}{\bibfnamefont{E.~A.} \bibnamefont{Jagla}},
  \bibinfo{journal}{J. Chem. Phys.} \textbf{\bibinfo{volume}{111}},
  \bibinfo{pages}{8980} (\bibinfo{year}{1999}{\natexlab{b}}).

\bibitem[{\citenamefont{Yan et~al.}(2006)\citenamefont{Yan, Buldyrev,
  Giovambattista, Debenedetti, and Stanley}}]{Yan06}
\bibinfo{author}{\bibfnamefont{Z.}~\bibnamefont{Yan}},
  \bibinfo{author}{\bibfnamefont{S.~V.} \bibnamefont{Buldyrev}},
  \bibinfo{author}{\bibfnamefont{N.}~\bibnamefont{Giovambattista}},
  \bibinfo{author}{\bibfnamefont{P.~G.} \bibnamefont{Debenedetti}},
  \bibnamefont{and} \bibinfo{author}{\bibfnamefont{H.~E.}
  \bibnamefont{Stanley}}, \bibinfo{journal}{Phys. Rev. E}
  \textbf{\bibinfo{volume}{73}}, \bibinfo{pages}{051204}
  (\bibinfo{year}{2006}).

\bibitem[{\citenamefont{Kumar et~al.}(2005)\citenamefont{Kumar, Buldyrev,
  Sciortino, Zaccarelli, and Stanley}}]{Kum05}
\bibinfo{author}{\bibfnamefont{P.}~\bibnamefont{Kumar}},
  \bibinfo{author}{\bibfnamefont{S.~V.} \bibnamefont{Buldyrev}},
  \bibinfo{author}{\bibfnamefont{F.}~\bibnamefont{Sciortino}},
  \bibinfo{author}{\bibfnamefont{E.}~\bibnamefont{Zaccarelli}},
  \bibnamefont{and} \bibinfo{author}{\bibfnamefont{H.~E.}
  \bibnamefont{Stanley}}, \bibinfo{journal}{Phys. Rev. E}
  \textbf{\bibinfo{volume}{72}}, \bibinfo{pages}{021501}
  (\bibinfo{year}{2005}).

\bibitem[{\citenamefont{Yan et~al.}(2005)\citenamefont{Yan, Buldyrev,
  Giovambattista, and Stanley}}]{Yan05}
\bibinfo{author}{\bibfnamefont{Z.}~\bibnamefont{Yan}},
  \bibinfo{author}{\bibfnamefont{S.~V.} \bibnamefont{Buldyrev}},
  \bibinfo{author}{\bibfnamefont{N.}~\bibnamefont{Giovambattista}},
  \bibnamefont{and} \bibinfo{author}{\bibfnamefont{H.~E.}
  \bibnamefont{Stanley}}, \bibinfo{journal}{Phys. Rev. Lett.}
  \textbf{\bibinfo{volume}{95}}, \bibinfo{pages}{130604}
  (\bibinfo{year}{2005}).

\bibitem[{\citenamefont{Malescio and Pellicane}(2003)}]{Mal03}
\bibinfo{author}{\bibfnamefont{G.}~\bibnamefont{Malescio}} \bibnamefont{and}
  \bibinfo{author}{\bibfnamefont{G.}~\bibnamefont{Pellicane}},
  \bibinfo{journal}{Nat. Mater.} \textbf{\bibinfo{volume}{2}},
  \bibinfo{pages}{97} (\bibinfo{year}{2003}).

\bibitem[{\citenamefont{Malescio and Pellicane}(2004)}]{Mal04}
\bibinfo{author}{\bibfnamefont{G.}~\bibnamefont{Malescio}} \bibnamefont{and}
  \bibinfo{author}{\bibfnamefont{G.}~\bibnamefont{Pellicane}},
  \bibinfo{journal}{Phys. Rev. E} \textbf{\bibinfo{volume}{70}},
  \bibinfo{pages}{021202} (\bibinfo{year}{2004}).

\bibitem[{\citenamefont{Glaser et~al.}(2007)\citenamefont{Glaser, Grason,
  Kamien, Ko$\mathrm{\check s}$mrlj, Santangelo, and Ziherl}}]{Gla07}
\bibinfo{author}{\bibfnamefont{M.}~\bibnamefont{Glaser}},
  \bibinfo{author}{\bibfnamefont{G.}~\bibnamefont{Grason}},
  \bibinfo{author}{\bibfnamefont{R.}~\bibnamefont{Kamien}},
  \bibinfo{author}{\bibfnamefont{A.}~\bibnamefont{Ko$\mathrm{\check s}$mrlj}},
  \bibinfo{author}{\bibfnamefont{C.}~\bibnamefont{Santangelo}},
  \bibnamefont{and} \bibinfo{author}{\bibfnamefont{P.}~\bibnamefont{Ziherl}},
  \bibinfo{journal}{Europhys. Lett.} \textbf{\bibinfo{volume}{78}},
  \bibinfo{pages}{46004} (\bibinfo{year}{2007}).

\bibitem[{\citenamefont{Fornleitner and Kahl}(2008)}]{For08Kah}
\bibinfo{author}{\bibfnamefont{J.}~\bibnamefont{Fornleitner}} \bibnamefont{and}
  \bibinfo{author}{\bibfnamefont{G.}~\bibnamefont{Kahl}},
  \bibinfo{journal}{Europhys. Lett.} \textbf{\bibinfo{volume}{82}},
  \bibinfo{pages}{18001} (\bibinfo{year}{2008}).

\bibitem[{\citenamefont{Pauschenwein and Kahl}(2008{\natexlab{a}})}]{Pau08}
\bibinfo{author}{\bibfnamefont{G.~J.} \bibnamefont{Pauschenwein}}
  \bibnamefont{and} \bibinfo{author}{\bibfnamefont{G.}~\bibnamefont{Kahl}},
  \bibinfo{journal}{Soft Matter} \textbf{\bibinfo{volume}{4}},
  \bibinfo{pages}{1396} (\bibinfo{year}{2008}{\natexlab{a}}).

\bibitem[{\citenamefont{Ziherl and Kamien}(2001)}]{Zih01}
\bibinfo{author}{\bibfnamefont{P.}~\bibnamefont{Ziherl}} \bibnamefont{and}
  \bibinfo{author}{\bibfnamefont{R.}~\bibnamefont{Kamien}},
  \bibinfo{journal}{J. Phys. Chem. B} \textbf{\bibinfo{volume}{105}},
  \bibinfo{pages}{10147} (\bibinfo{year}{2001}).

\bibitem[{\citenamefont{Norizoe and Kawakatsu}(2005)}]{Nor05}
\bibinfo{author}{\bibfnamefont{Y.}~\bibnamefont{Norizoe}} \bibnamefont{and}
  \bibinfo{author}{\bibfnamefont{T.}~\bibnamefont{Kawakatsu}},
  \bibinfo{journal}{Europhys. Lett.} \textbf{\bibinfo{volume}{72}},
  \bibinfo{pages}{583} (\bibinfo{year}{2005}).

\bibitem[{\citenamefont{Pierleoni et~al.}(2006)\citenamefont{Pierleoni,
  Addison, Hansen, and Krakoviack}}]{Pie06}
\bibinfo{author}{\bibfnamefont{C.}~\bibnamefont{Pierleoni}},
  \bibinfo{author}{\bibfnamefont{C.}~\bibnamefont{Addison}},
  \bibinfo{author}{\bibfnamefont{J.-P.} \bibnamefont{Hansen}},
  \bibnamefont{and}
  \bibinfo{author}{\bibfnamefont{V.}~\bibnamefont{Krakoviack}},
  \bibinfo{journal}{Phys. Rev. Lett.} \textbf{\bibinfo{volume}{96}},
  \bibinfo{pages}{128302} (\bibinfo{year}{2006}).

\bibitem[{\citenamefont{Campbell et~al.}(2005)\citenamefont{Campbell, Anderson,
  van Duijneveldt, and Bartlett}}]{Cam05}
\bibinfo{author}{\bibfnamefont{A.}~\bibnamefont{Campbell}},
  \bibinfo{author}{\bibfnamefont{V.}~\bibnamefont{Anderson}},
  \bibinfo{author}{\bibfnamefont{J.}~\bibnamefont{van Duijneveldt}},
  \bibnamefont{and} \bibinfo{author}{\bibfnamefont{P.}~\bibnamefont{Bartlett}},
  \bibinfo{journal}{Phys. Rev. Lett.} \textbf{\bibinfo{volume}{94}},
  \bibinfo{pages}{208301} (\bibinfo{year}{2005}).

\bibitem[{\citenamefont{Camp}(2003)}]{Cam03}
\bibinfo{author}{\bibfnamefont{P.}~\bibnamefont{Camp}}, \bibinfo{journal}{Phys.
  Rev. E} \textbf{\bibinfo{volume}{68}}, \bibinfo{pages}{061506}
  (\bibinfo{year}{2003}).

\bibitem[{\citenamefont{de~Candia et~al.}(2006)\citenamefont{de~Candia, Gado,
  Fierro, Sator, Tarzia, and Coniglio}}]{Can06}
\bibinfo{author}{\bibfnamefont{A.}~\bibnamefont{de~Candia}},
  \bibinfo{author}{\bibfnamefont{E.~D.} \bibnamefont{Gado}},
  \bibinfo{author}{\bibfnamefont{A.}~\bibnamefont{Fierro}},
  \bibinfo{author}{\bibfnamefont{N.}~\bibnamefont{Sator}},
  \bibinfo{author}{\bibfnamefont{M.}~\bibnamefont{Tarzia}}, \bibnamefont{and}
  \bibinfo{author}{\bibfnamefont{A.}~\bibnamefont{Coniglio}},
  \bibinfo{journal}{Phys. Rev. E} \textbf{\bibinfo{volume}{74}},
  \bibinfo{pages}{010403(R)} (\bibinfo{year}{2006}).

\bibitem[{\citenamefont{Stradner et~al.}(2004)\citenamefont{Stradner, Sedgwick,
  Cardinaux, Poon, Egelhaaf, and Schurtenberger}}]{Str04}
\bibinfo{author}{\bibfnamefont{A.}~\bibnamefont{Stradner}},
  \bibinfo{author}{\bibfnamefont{H.}~\bibnamefont{Sedgwick}},
  \bibinfo{author}{\bibfnamefont{F.}~\bibnamefont{Cardinaux}},
  \bibinfo{author}{\bibfnamefont{W.~C.~K.} \bibnamefont{Poon}},
  \bibinfo{author}{\bibfnamefont{S.~U.} \bibnamefont{Egelhaaf}},
  \bibnamefont{and}
  \bibinfo{author}{\bibfnamefont{P.}~\bibnamefont{Schurtenberger}},
  \bibinfo{journal}{Nature} \textbf{\bibinfo{volume}{432}},
  \bibinfo{pages}{492} (\bibinfo{year}{2004}).

\bibitem[{\citenamefont{Mladek et~al.}(2006{\natexlab{a}})\citenamefont{Mladek,
  Gottwald, Kahl, Neumann, and Likos}}]{Mla06PRL}
\bibinfo{author}{\bibfnamefont{B.}~\bibnamefont{Mladek}},
  \bibinfo{author}{\bibfnamefont{D.}~\bibnamefont{Gottwald}},
  \bibinfo{author}{\bibfnamefont{G.}~\bibnamefont{Kahl}},
  \bibinfo{author}{\bibfnamefont{M.}~\bibnamefont{Neumann}}, \bibnamefont{and}
  \bibinfo{author}{\bibfnamefont{C.}~\bibnamefont{Likos}},
  \bibinfo{journal}{Phys. Rev. Lett.} \textbf{\bibinfo{volume}{96}},
  \bibinfo{pages}{045701} (\bibinfo{year}{2006}{\natexlab{a}}).

\bibitem[{\citenamefont{Mladek et~al.}(2006{\natexlab{b}})\citenamefont{Mladek,
  Gottwald, Kahl, Neumann, and Likos}}]{Mla06PRLERR}
\bibinfo{author}{\bibfnamefont{B.}~\bibnamefont{Mladek}},
  \bibinfo{author}{\bibfnamefont{D.}~\bibnamefont{Gottwald}},
  \bibinfo{author}{\bibfnamefont{G.}~\bibnamefont{Kahl}},
  \bibinfo{author}{\bibfnamefont{M.}~\bibnamefont{Neumann}}, \bibnamefont{and}
  \bibinfo{author}{\bibfnamefont{C.}~\bibnamefont{Likos}},
  \bibinfo{journal}{Phys. Rev. Lett.} \textbf{\bibinfo{volume}{97}},
  \bibinfo{pages}{019901} (\bibinfo{year}{2006}{\natexlab{b}});
  {\it erratum} to \cite{Mla06PRL}.

\bibitem[{\citenamefont{Holland}(1975)}]{Hol75}
\bibinfo{author}{\bibfnamefont{J.}~\bibnamefont{Holland}},
  \emph{\bibinfo{title}{Adaption in Natural and Artificial Systems}}
  (\bibinfo{publisher}{The University of Michigan Press}, \bibinfo{address}{Ann
  Arbor}, \bibinfo{year}{1975}).

\bibitem[{\citenamefont{Woodley et~al.}(1999)\citenamefont{Woodley, Battle,
  Gale, and Catlow}}]{Woo99}
\bibinfo{author}{\bibfnamefont{S.~M.} \bibnamefont{Woodley}},
  \bibinfo{author}{\bibfnamefont{P.~D.} \bibnamefont{Battle}},
  \bibinfo{author}{\bibfnamefont{J.~D.} \bibnamefont{Gale}}, \bibnamefont{and}
  \bibinfo{author}{\bibfnamefont{C.~R.~A.} \bibnamefont{Catlow}},
  \bibinfo{journal}{Phys. Chem. Chem. Phys.} \textbf{\bibinfo{volume}{1}},
  \bibinfo{pages}{2535} (\bibinfo{year}{1999}).

\bibitem[{\citenamefont{Oganov and Glass}(2006)}]{Oga06}
\bibinfo{author}{\bibfnamefont{A.~R.} \bibnamefont{Oganov}} \bibnamefont{and}
  \bibinfo{author}{\bibfnamefont{C.~W.} \bibnamefont{Glass}},
  \bibinfo{journal}{J. Chem. Phys.} \textbf{\bibinfo{volume}{124}},
  \bibinfo{pages}{244704} (\bibinfo{year}{2006}).

\bibitem[{\citenamefont{Oganov and Glass}(2008)}]{Oga08}
\bibinfo{author}{\bibfnamefont{A.~R.} \bibnamefont{Oganov}} \bibnamefont{and}
  \bibinfo{author}{\bibfnamefont{C.~W.} \bibnamefont{Glass}},
  \bibinfo{journal}{J. Phys.: Condens. Mat.} \textbf{\bibinfo{volume}{20}},
  \bibinfo{pages}{064210} (\bibinfo{year}{2008}).

\bibitem[{\citenamefont{Gottwald et~al.}(2004)\citenamefont{Gottwald, Likos,
  Kahl, and L{\"o}wen}}]{Got04}
\bibinfo{author}{\bibfnamefont{D.}~\bibnamefont{Gottwald}},
  \bibinfo{author}{\bibfnamefont{C.}~\bibnamefont{Likos}},
  \bibinfo{author}{\bibfnamefont{G.}~\bibnamefont{Kahl}}, \bibnamefont{and}
  \bibinfo{author}{\bibfnamefont{H.}~\bibnamefont{L{\"o}wen}},
  \bibinfo{journal}{Phys. Rev. Lett.} \textbf{\bibinfo{volume}{92}},
  \bibinfo{pages}{068301} (\bibinfo{year}{2004}).

\bibitem[{\citenamefont{Gottwald
  et~al.}(2005{\natexlab{a}})\citenamefont{Gottwald, Kahl, and
  Likos}}]{Got05Kah}
\bibinfo{author}{\bibfnamefont{D.}~\bibnamefont{Gottwald}},
  \bibinfo{author}{\bibfnamefont{G.}~\bibnamefont{Kahl}}, \bibnamefont{and}
  \bibinfo{author}{\bibfnamefont{C.}~\bibnamefont{Likos}}, \bibinfo{journal}{J.
  Chem. Phys.} \textbf{\bibinfo{volume}{122}}, \bibinfo{pages}{204503}
  (\bibinfo{year}{2005}{\natexlab{a}}).

\bibitem[{\citenamefont{Gottwald
  et~al.}(2005{\natexlab{b}})\citenamefont{Gottwald, Likos, Kahl, and
  L{\"o}wen}}]{Got05Lik}
\bibinfo{author}{\bibfnamefont{D.}~\bibnamefont{Gottwald}},
  \bibinfo{author}{\bibfnamefont{C.}~\bibnamefont{Likos}},
  \bibinfo{author}{\bibfnamefont{G.}~\bibnamefont{Kahl}}, \bibnamefont{and}
  \bibinfo{author}{\bibfnamefont{H.}~\bibnamefont{L{\"o}wen}},
  \bibinfo{journal}{J. Chem. Phys.} \textbf{\bibinfo{volume}{122}},
  \bibinfo{pages}{074903} (\bibinfo{year}{2005}{\natexlab{b}}).

\bibitem[{\citenamefont{Fornleitner et~al.}(2008)\citenamefont{Fornleitner,
  \mbox{Lo Verso}, Kahl, and \mbox{C. N}. Likos}}]{For08LoV}
\bibinfo{author}{\bibfnamefont{J.}~\bibnamefont{Fornleitner}},
  \bibinfo{author}{\bibfnamefont{F.}~\bibnamefont{\mbox{Lo Verso}}},
  \bibinfo{author}{\bibfnamefont{G.}~\bibnamefont{Kahl}}, \bibnamefont{and}
  \bibinfo{author}{\bibnamefont{\mbox{C. N}. Likos}}, \bibinfo{journal}{Soft
  Matter} \textbf{\bibinfo{volume}{4}}, \bibinfo{pages}{480}
  (\bibinfo{year}{2008}).

\bibitem[{\citenamefont{Pauschenwein and
  Kahl}(2008{\natexlab{b}})}]{Pau08mindist}
\bibinfo{author}{\bibfnamefont{G.~J.} \bibnamefont{Pauschenwein}}
  \bibnamefont{and} \bibinfo{author}{\bibfnamefont{G.}~\bibnamefont{Kahl}},
  \bibinfo{journal}{to be published}  (\bibinfo{year}{2008}{\natexlab{b}}).

\bibitem[{\citenamefont{Powell}(1964)}]{Pow64}
\bibinfo{author}{\bibfnamefont{M.~J.~D.} \bibnamefont{Powell}},
  \bibinfo{journal}{Comput. J.} \textbf{\bibinfo{volume}{7}},
  \bibinfo{pages}{152} (\bibinfo{year}{1964}).

\bibitem{rareEarth} Note that the formation of non-trivial
  close-packed arrangements is also reported for certain rare earth
  elements (page 79 in \cite {Ash76}).

\bibitem[{\citenamefont{Ashcroft and Mermin}(1976)}]{Ash76}
\bibinfo{author}{\bibfnamefont{N.~W.} \bibnamefont{Ashcroft}} \bibnamefont{and}
  \bibinfo{author}{\bibfnamefont{N.~D.} \bibnamefont{Mermin}},
  \emph{\bibinfo{title}{Solid State Physics}} (\bibinfo{publisher}{Saunders
  College Publishing, Harcourt Brace College Publishers},
  \bibinfo{address}{Fort Worth, Philadelphia, San Diego, New York, Orlando,
  Austin, San Antonio, Toronto, Montreal, London, Sydney, Tokyo},
  \bibinfo{year}{1976}).

\bibitem[{\citenamefont{Pauschenwein}(2008)}]{Pau08th}
\bibinfo{author}{\bibfnamefont{G.~J.} \bibnamefont{Pauschenwein}}, Ph.D.
  thesis, \bibinfo{school}{Institut f{\"u}r Theoretische Physik, TU Wien}
  (\bibinfo{year}{2008}).

\bibitem{fccFavor} This choice is motivated by the fact that --
  although the stacking sequence for fcc is longer than the one for
  hcp -- the crystallographic description for the fcc requires only
  one basis particle, while for the hcp structure it is a non-simple
  one.

\bibitem[{\citenamefont{\mbox{C. N.} Likos et~al.}(1998)\citenamefont{\mbox{C.
  N.} Likos, Watzlawek, and L{\"o}wen}}]{Lik98Wat}
\bibinfo{author}{\bibnamefont{\mbox{C. N.} Likos}},
  \bibinfo{author}{\bibfnamefont{M.}~\bibnamefont{Watzlawek}},
  \bibnamefont{and}
  \bibinfo{author}{\bibfnamefont{H.}~\bibnamefont{L{\"o}wen}},
  \bibinfo{journal}{Phys. Rev. E} \textbf{\bibinfo{volume}{58}},
  \bibinfo{pages}{3135} (\bibinfo{year}{1998}).

\bibitem[{\citenamefont{Likos et~al.}(2001)\citenamefont{Likos, Lang,
  Watzlawek, and L{\"o}wen}}]{Lik01Lan}
\bibinfo{author}{\bibfnamefont{C.~N.} \bibnamefont{Likos}},
  \bibinfo{author}{\bibfnamefont{A.}~\bibnamefont{Lang}},
  \bibinfo{author}{\bibfnamefont{M.}~\bibnamefont{Watzlawek}},
  \bibnamefont{and}
  \bibinfo{author}{\bibfnamefont{H.}~\bibnamefont{L{\"o}wen}},
  \bibinfo{journal}{Phys. Rev. E} \textbf{\bibinfo{volume}{63}},
  \bibinfo{pages}{31206} (\bibinfo{year}{2001}).

\bibitem[{\citenamefont{Falkinger et~al.}(2008)\citenamefont{Falkinger, Mladek,
  Gottwald, and Kahl}}]{Fal08}
\bibinfo{author}{\bibfnamefont{G.}~\bibnamefont{Falkinger}},
  \bibinfo{author}{\bibfnamefont{B.}~\bibnamefont{Mladek}},
  \bibinfo{author}{\bibfnamefont{D.}~\bibnamefont{Gottwald}}, \bibnamefont{and}
  \bibinfo{author}{\bibfnamefont{G.}~\bibnamefont{Kahl}},
  \bibinfo{journal}{submitted to J. Phys.: Condens. Mat.}
  (\bibinfo{year}{2008}).

\bibitem[{\citenamefont{\mbox{C. N.} Likos et~al.}(2007)\citenamefont{\mbox{C.
  N.} Likos, \mbox{B. M.} Mladek, Gottwald, and Kahl}}]{Lik07}
\bibinfo{author}{\bibnamefont{\mbox{C. N.} Likos}},
  \bibinfo{author}{\bibnamefont{\mbox{B. M.} Mladek}},
  \bibinfo{author}{\bibfnamefont{D.}~\bibnamefont{Gottwald}}, \bibnamefont{and}
  \bibinfo{author}{\bibfnamefont{G.}~\bibnamefont{Kahl}}, \bibinfo{journal}{J.
  Chem. Phys.} \textbf{\bibinfo{volume}{126}}, \bibinfo{pages}{224502}
  (\bibinfo{year}{2007}).

\bibitem[{\citenamefont{Mladek et~al.}(2008)\citenamefont{Mladek, Gottwald,
  Kahl, Neumann, and Likos}}]{Mla07Got}
\bibinfo{author}{\bibfnamefont{B.~M.} \bibnamefont{Mladek}},
  \bibinfo{author}{\bibfnamefont{D.}~\bibnamefont{Gottwald}},
  \bibinfo{author}{\bibfnamefont{G.}~\bibnamefont{Kahl}},
  \bibinfo{author}{\bibfnamefont{M.}~\bibnamefont{Neumann}}, \bibnamefont{and}
  \bibinfo{author}{\bibfnamefont{C.~N.} \bibnamefont{Likos}},
  \bibinfo{journal}{J. Phys. Chem. B} \textbf{\bibinfo{volume}{111}},
  \bibinfo{pages}{12799} (\bibinfo{year}{2008}).

\bibitem[{\citenamefont{Evans}(1992)}]{Eva92}
\bibinfo{author}{\bibfnamefont{R.}~\bibnamefont{Evans}},
  \emph{\bibinfo{title}{Fundamentals of Inhomogeneous Fluids}}
  (\bibinfo{publisher}{Marcel Dekker, New York}, \bibinfo{year}{1992}),
  chap.~\bibinfo{chapter}{3}, pp. \bibinfo{pages}{85--175}.

\bibitem[{\citenamefont{Roth et~al.}(2002)\citenamefont{Roth, Evans, Lang, and
  Kahl}}]{Rot02}
\bibinfo{author}{\bibfnamefont{R.}~\bibnamefont{Roth}},
  \bibinfo{author}{\bibfnamefont{R.}~\bibnamefont{Evans}},
  \bibinfo{author}{\bibfnamefont{A.}~\bibnamefont{Lang}}, \bibnamefont{and}
  \bibinfo{author}{\bibfnamefont{G.}~\bibnamefont{Kahl}}, \bibinfo{journal}{J.
  Phys.: Condens. Mat.} \textbf{\bibinfo{volume}{14}}, \bibinfo{pages}{12063}
  (\bibinfo{year}{2002}).

\bibitem[{\citenamefont{Fomin et~al.}(2006)\citenamefont{Fomin, Gribova,
  Ryzhov, Frenkel, and Stishov}}]{Fom06}
\bibinfo{author}{\bibfnamefont{Y.~D.} \bibnamefont{Fomin}},
  \bibinfo{author}{\bibfnamefont{N.~V.} \bibnamefont{Gribova}},
  \bibinfo{author}{\bibfnamefont{V.~N.} \bibnamefont{Ryzhov}},
  \bibinfo{author}{\bibfnamefont{D.}~\bibnamefont{Frenkel}}, \bibnamefont{and}
  \bibinfo{author}{\bibfnamefont{S.~M.} \bibnamefont{Stishov}},
  \bibinfo{journal}{ArXiv:cond-mat} \textbf{\bibinfo{volume}{612586}}
  (\bibinfo{year}{2006}).
\end{thebibliography}

\eject
\section{Tables}
\begin{table}[!htb]
\caption{\label{tab:abbrev} Standard abbreviations for the 14 Bravais
  lattices used in the text and the captions. The seven crystal
  systems are separated by double lines.}
\end{table}

\renewcommand{\arraystretch}{0.6}
\begin{tabular}{l|c}
  \multicolumn{1}{c|}{Bravais lattice} & abbreviation \\\hline\hline
  simple cubic & sc \\\hline
  body centered cubic & bcc \\\hline
  face centered cubic & fcc \\\hline\hline
  hexagonal & hex \\\hline\hline
  trigonal (rhombohedral) & trig \\\hline\hline
  simple tetragonal & st \\\hline
  centered tetragonal & ct \\\hline\hline
  simple orthorhombic & so \\\hline
  single face centered orthorhombic & sfco \\\hline
  body centered orthorhombic & bco \\\hline
  face centered orthorhombic & fco \\\hline\hline
  simple monoclinic & sm \\\hline
  single face centered monoclinic & sfcm \\\hline\hline
  triclinic & tric 
\end{tabular}
\begin{table}[!htb]
\caption{\label{tab:1.5} Numerical details of the ordered equilibrium
  structures identified for the square-shoulder system with
  $\lambda=1.5$: the underlying lattice is characterized by the
  according abbreviation (cf.~Table \ref{tab:abbrev}), $n_b$ is the
  number of basis particles required to describe the MEC. $e^\star$
  and $\rho^\star$ are the energy per particle and particle density,
  respectively. Since the MECs can be interpreted on the basis of
  geometric considerations, $\rho^\star$ can be given in closed,
  analytic expressions.}
\end{table}

\vspace{-5mm}
\renewcommand{\arraystretch}{1}
\[\begin{array}{l|c|c|c@{\;\simeq\;}l}
\mbox{lattice}& n_b &e^\star &\multicolumn{2}{c}{ \rho^\star}\\\hline\hline 
\mbox{fcc}&1& 0 & \frac{8\,{\sqrt{2}}}{27} & 0.419\\\hline
\mbox{bco}&1& 1 & \frac{8}{3\,{\sqrt{23}}} & 0.556 \\\hline 
\mbox{fco}&2& 2 & \frac{16}{{\sqrt{455}}} & 0.750\\\hline 
\mbox{hex}&1& 4 & \frac{4}{{\sqrt{15}}} & 1.03\\\hline
\mbox{sfco}&2& 5 & \frac{16}{2\,{\sqrt{15}} + {\sqrt{35}}} & 1.17\\\hline
\mbox{bco} & 1& 7 & \frac{8}{{\sqrt{35}}} & 1.35\\\hline
\mbox{fcc}&1& 9 & \sqrt{2} & 1.41
\end{array}\]
\\\\\\\\\\\\\\\\\\

\begin{table}[!htb]
\caption{\label{tab:4.5_array} Numerical details of the ordered
  equilibrium structures identified for the square-shoulder system
  with $\lambda=4.5$: the underlying lattice is characterized by the
  according abbreviation (cf.~Table \ref{tab:abbrev}), $n_b$ is the
  number of basis particles required to describe the MEC. $e^\star$
  and $\rho^\star$ are the energy per particle and particle density,
  respectively. The abbreviation in the third row indicates to which
  of the four archetypes the MEC belongs (\emph{clu}ster,
  \emph{col}umnar, \emph{lam}ellar, \emph{com}pact). Closed, algebraic
  expressions for $\rho^\star$ can be derived, but are not presented
  here due to space limitations. For details cf.~\cite{Pau08th}.}
\end{table}
\vspace{-4cm}
\renewcommand{\arraystretch}{0.5}
\[\begin{array}{l|c|c|c|r@{.}l}
\mbox{lattice}& n_b & \mbox{shape} &e^\star &
\multicolumn{2}{c}{\rho^\star} \\\hline\hline 
\mbox{fcc} & 1 & \mbox{clu} & 0 & 0&0155 \\\hline
\mbox{sm} & 2 & \mbox{clu} & 1/2 & 0&0235\\\hline
\mbox{sfcm} & 3 & \mbox{clu} & 1 & 0&0291  \\\hline
\mbox{tric} & 8 & \mbox{clu (4)} & 3/2 & 0&0343\\\hline
\mbox{tric} & 4 & \mbox{clu} & 7/4 & 0&0369\\\hline
\mbox{tric} & 6 & \mbox{clu} & 5/2 & 0&0454\\\hline
\mbox{bco} & 1 & \mbox{col} & 3 & 0&0512  \\\hline
\mbox{bco} & 8 & \mbox{clu} & {35}/{8} & 0&0638\\\hline
\mbox{tric} & 2 & \mbox{col} & {13}/{2} & 0&0799\\\hline
\mbox{sm} & 2 & \mbox{col} & 8 & 0&0923  \\\hline 
\mbox{tric} & 2 & \mbox{col} & {21}/{2} & 0&104\\\hline
\mbox{bco} & 1 & \mbox{lam/col} & 15 & 0&135  \\\hline
\mbox{trig} & 1 & \mbox{lam} & 18 & 0&154 \\\hline
\mbox{tric} & 2 & \mbox{lam} & 20 & 0&168 \\\hline
\mbox{tric} & 2 & \mbox{lam (col)} & {49}/{2} & 0&198 \\\hline
\mbox{trig} & 2 & \mbox{lam} & 30 & 0&243 \\\hline
\mbox{trig} & 2 & \mbox{lam} & {61}/{2} & 0&245\\\hline
\mbox{ct} & 2 & \mbox{lam} & {77}/{2} & 0&283\\\hline
\mbox{sfcm} & 2 & \mbox{lam} & 56 & 0&383  \\\hline
\mbox{sm} & 2 & \mbox{lam} & 58 & 0&394  \\\hline
\mbox{fco} & 2 & \mbox{lam} & {143}/{2} & 0&484\\\hline
\mbox{bcc} & 1 & \mbox{com} & 90 & 0&567  \\\hline
\mbox{ct} & 1 & \mbox{com} & 96 & 0&594   \\\hline
\mbox{tric} & 3 & \mbox{lam} & {337}/{3} & 0&677\\\hline
\mbox{trig} & 1 & \mbox{com} & 115 & 0&692  \\\hline
\mbox{sc} & 1 & \mbox{com} & 128 & 0&763   \\\hline
\mbox{hex} & 1 & \mbox{com} & 138 & 0&811  \\\hline
\mbox{bcc} & 1 & \mbox{com} & 169 & 0&997  \\\hline
\mbox{ct} & 1 & \mbox{com} & 180 & 1&05   \\\hline
\mbox{sfcm} & 1 & \mbox{com} & 210 & 1&21  \\\hline
\mbox{fcc} & 1 & \mbox{com} & 229 & 1&29   \\\hline
\mbox{ct} & 1 & \mbox{com} & 243 & 1&34   \\\hline
\mbox{hcp} & 2 & \mbox{com} & 263 & 1&41 
\end{array}\]

\end{document}